\documentclass[12pt]{article}
\pdfoutput=1
\usepackage[nosort]{cite}

\usepackage{draft,hyperref,tikz,float,subfig,cite,ifthen}

\usetikzlibrary{decorations.markings}
\tikzset{middlearrow/.style={
	decoration={markings,
		mark=at position 0.55 with {\arrow[>=stealth]{#1}}},
	postaction={decorate}
	}
}

\tikzset{line/.style={line width=0.5mm},
curve/.style={line,smooth,tension=1},
->-/.style={decoration={
  markings,
  mark=at position #1 with {\arrow[>=stealth]{>}}},postaction={decorate}},
-<-/.style={decoration={
  markings,
  mark=at position #1 with {\arrow[>=stealth]{<}}},postaction={decorate}},
}

\usetikzlibrary{snakes}
\usetikzlibrary{shapes.misc}

\newcommand{\Zpre}[5]{
\tikzset{line/.style={line width=0.25mm},
curve/.style={line,smooth,tension=1}}
\draw [line] (0,0) -- (2,0) -- (2,2) -- (0,2) -- (0,0);
\ifthenelse{\equal{#4}{}}{
	\ifthenelse{\equal{#2}{1}}{\draw [line,dashed] (0,1) -- (2,1)}{
		\ifthenelse{\equal{#2}{2}}{\draw [line,middlearrow={>},line width=0.5mm] (0,1) -- (2,1)}{}
	};
	\ifthenelse{\equal{#3}{1}}{\draw [line,dashed] (1,0) -- (1,2)}{
		\ifthenelse{\equal{#3}{2}}{\draw [line,middlearrow={>},line width=0.5mm] (1,0) -- (1,2)}{}
	};
}{
	\ifthenelse{\equal{#3}{1}}{
			\draw [line,dashed] (1,0) -- (1,0.7);
			\draw [line,dashed] (1,2) -- (1,1.3);
			\draw [line] (1,0.7) -- (1,1.3);
			
		}{
		\ifthenelse{\equal{#3}{2} \AND \not\equal{#2}{2}}{\draw [line] (1,0) -- (1,2)}{
			\ifthenelse{\equal{#3}{2} \AND \equal{#2}{2}}{
				\draw [line] (1,0) -- (1,0.7);
				\draw [line] (1,2) -- (1,1.3);
				\ifthenelse{\equal{#5}{1}}{
					\draw [line,dashed] (1,0.7) -- (1,1.3);
				}{};
			}{};
		};
	};
	\ifthenelse{\equal{#4}{1}}{
		\ifthenelse{\equal{#2}{1}}{
			\draw [curve,dashed] plot coordinates {(0,1) (0.7,1) (1,1.3)};
			\draw [curve,dashed] plot coordinates {(2,1) (1.3,1) (1,0.7)};
			}{
			\ifthenelse{\equal{#2}{2}}{
			\draw [curve] plot coordinates {(0,1) (0.7,1) (1,1.3)};
			\draw [curve] plot coordinates {(2,1) (1.3,1) (1,0.7)};
			}{};
		};
	}{
		\ifthenelse{\equal{#4}{0}}{
			\ifthenelse{\equal{#2}{1}}{
				\draw [curve,dashed] plot coordinates {(0,1) (0.7,1) (1,0.7)};
				\draw [curve,dashed] plot coordinates {(2,1) (1.3,1) (1,1.3)};
				}{
				\ifthenelse{\equal{#2}{2}}{
				\draw [curve] plot coordinates {(0,1) (0.7,1) (1,0.7)};
				\draw [curve] plot coordinates {(2,1) (1.3,1) (1,1.3)};
				}{};
			};
		};
	};
};
}

\newcommand{\Z}[5]{
\begin{gathered}
\begin{tikzpicture}[scale = #1]
\Zpre{#1}{#2}{#3}{#4}{#5}
\end{tikzpicture}
\end{gathered}
}

\begin{document}

\begin{titlepage}

\preprint{CALT-TH-2021-002}

\begin{center}

\hfill \\
\hfill \\
\vskip 1cm

\title{
$\mathbb{Z}_N$ Symmetries,  Anomalies, and the Modular Bootstrap
}

\author{Ying-Hsuan Lin$^{a,b}$ and Shu-Heng Shao$^c$
}

\address{${}^a$Jefferson Physical Laboratory, Harvard University, Cambridge, MA 02138, USA}
\address{${}^b$Walter Burke Institute for Theoretical Physics,\\ California Institute of Technology,
Pasadena, CA 91125, USA}
\address{${}^c$School of Natural Sciences, Institute for Advanced Study,\\
Princeton, NJ 08540, USA}

\email{yhlin@fas.harvard.edu, shao@ias.edu}

\end{center}

\vfill

\abstract{
We explore constraints on  (1+1)$d$ unitary conformal field theory with an internal $\bZ_N$ global symmetry, by bounding the lightest symmetry-preserving scalar primary operator using the modular bootstrap.   
Among the other constraints we have found, we prove the existence of a $\bZ_N$-symmetric relevant/marginal operator if $N-1 \le c\le 9-N$ for $N\leq4$, with the endpoints  saturated by various WZW models that can be embedded into $(\mathfrak{e}_8)_1$. 
Its existence implies that robust gapless fixed points are not possible in this range of $c$  if only a $\bZ_N$ symmetry is imposed microscopically.
We also obtain stronger, more refined bounds that depend on the 't Hooft anomaly of the $\bZ_N$ symmetry. 
}

\vfill

\end{titlepage}

\eject
\setcounter{tocdepth}{2} 

\tableofcontents

\section{Introduction}

Global symmetries and their 't Hooft anomalies ({\it i.e.} obstruction to gauging) are central tools in analyzing strongly coupled quantum systems. 
In this paper, we continue our exploration in \cite{Lin:2019kpn} of  universal  constraints imposed by the symmetries and  anomalies in  conformal field theory (CFT).   
We will apply the techniques of the conformal bootstrap, which exploits the internal consistency of CFT, to derive general constraints on (1+1)$d$ unitary bosonic CFT with  $\mathbb{Z}_N$ global symmetry.

Of special importance are constraints on the robust (or stable),  gapless fixed points of renormalization group (RG) flows.  
Typically, we start from a short-distanced quantum system, which can for example be another quantum field theory or a lattice model, and impose a global symmetry $G_\text{UV}$ on the system.  
The global symmetry $G_\text{UV}$ generally carries  certain 't Hooft anomalies. 
We would like to ask whether this system can flow to a  gapless fixed point without fine-tuning the  parameters at short distances.
Typically the global symmetry in the infrared (IR) CFT is enhanced to a bigger group $G_\text{IR}$ with an embedding $G_\text{UV} \to G_\text{IR}$, in which case 
the IR CFT has to match the 't Hooft anomalies of the UV system.

A generic local  perturbation of the UV system becomes a local   operator in the low-energy quantum field theory.  
For the candidate low-energy CFT to be a robust fixed point, there   cannot be any $G_\text{UV}$-invariant relevant local  operator.\footnote{However, see \cite{Alavirad:2019iea} for an interesting example of a (1+1)$d$ spin chain where the low-energy gapless fixed point is stable even in the presence of symmetry-preserving relevant local operators. } 
The absence of such an operator at low energy guarantees that if the flow approaches close to the fixed point, it will be attracted to it.  
See, for example, \cite{Affleck:1988zj,Affleck:1988wz,Ohmori:2018qza} for applications of this type of argument, and \cite{Seiberg:2020bhn} for a recent discussion on the robustness of quantum field theory. 

Our approach to this general question is to exploit the internal consistency of CFT in (1+1) dimensions.  
In (1+1)$d$ CFT, the spectrum of local operators is highly constrained by the modular invariance of the torus partition function.  
Most famously, Cardy derived a universal formula for the density of heavy local operators in terms of the central charge \cite{Cardy:1986ie}. 
More generally, when the CFT has a certain global symmetry,  the spectra of charged and uncharged operators   are  constrained by the modular covariance of the torus partition function dressed with symmetry defects.  
The modern modular bootstrap program, which exploits the above consistency conditions, provides an ideal platform to map out the space of CFTs with global symmetry. 
See, for example,
\cite{Hellerman:2009bu,Hellerman:2010qd,Keller:2012mr,Friedan:2013cba,Qualls:2013eha,Hartman:2014oaa,Qualls:2014oea,Kim:2015oca,Benjamin:2016fhe,Collier:2016cls,Collier:2017shs,Cho:2017fzo,Keller:2017iql,Cardy:2017qhl,Bae:2017kcl,Dyer:2017rul,Anous:2018hjh,Bae:2018qym,Afkhami-Jeddi:2019zci,Mukhametzhanov:2019pzy,Lin:2019kpn,Hartman:2019pcd,Ganguly:2019ksp,Benjamin:2019stq,Pal:2019zzr,Benjamin:2020swg,Mukhametzhanov:2020swe,Pal:2020wwd,Benjamin:2020zbs,Afkhami-Jeddi:2020hde,Alday:2020qkm,Dymarsky:2020bps,Dymarsky:2020qom} for various exciting developments in the modular bootstrap.

In the special case of a $\bZ_N$ global symmetry with small $N$, we will derive a universal  upper bound $\Delta_\text{scalar}^{Q=0}$ on the scaling dimension of the lightest symmetry-preserving    scalar operator. 
This generalizes the previous works when there is no symmetry \cite{Collier:2016cls} and when the symmetry is $\bZ_2$ \cite{Lin:2019kpn}. 
This upper bound $\Delta_\text{scalar}^{Q=0}(N,k,c)$   depends on the central charge $c$ and the 't Hooft anomaly, denoted as $k$, of the $\bZ_N$ global symmetry.\footnote{For simplicity, we assume that the left and right central charges are equal in this paper, {\it i.e.} $c\equiv c_L=c_R $.}

For small $N$ and a given anomaly $k$, our bound further implies that there must be a symmetry-preserving relevant scalar operator  for any CFT within a certain range of the central charge.  
In other words, robust gapless fixed points are impossible within this range of the central charge if we only impose a $\bZ_N$ symmetry  microscopically.\footnote{The prototypical example of   a robust gapless phase with a non-anomalous $\bZ_N$ global symmetry is the $\bZ_N$ clock model \cite{PhysRevB.16.1217}.   
For $N\ge 5$, it is known that there is a region in the parameter space where the low energy phase is gapless  \cite{PhysRevD.19.3698,Cardy_1980,Alcaraz:1980sa}. 
Curiously, our numerical bootstrap bounds for the non-anomalous $\bZ_N$ symmetry  only give   a nontrivial range of $c$ when $N\leq 4$, and therefore do not constrain the  phase diagram of the clock model. We thank Pranay Gorantla, Ho Tat Lam, and Nathan Seiberg for related discussions.}

In (1+1)$d$, the anomaly of a global symmetry is encoded in the $F$-symbols of the fusion category characterizing the symmetry defects (see, for example, \cite{Bhardwaj:2017xup,Chang:2018iay}).  
For each $N$, there is a special value of the anomaly $k$ that is natural from the point of view of (2+1)$d$ topological quantum field theory (TQFT), which is described by unitary modular tensor category (UMTC). 
Specifically, for $k=0$ when  $N$ is odd and for $k=N/2$ when $N$ is even, the corresponding $\bZ_N$ fusion category with anomaly $k$ can be lifted to a UMTC  with $\bZ_N$ fusion rule \cite{Moore:1988qv} (see also \cite{Bonderson:2007ci,Barkeshli:2014cna,Hsin:2018vcg}).

For this special anomaly and small $N$, we find that the bounds at $c = N-1$ and $c= 9-N$ are saturated by the center symmetries of a pair of Wess-Zumino-Witten (WZW) models $(\mathfrak{g}_1 , \mathfrak{h}_1)$ at level 1. In these cases, the marginal operators are the current bilinear operators $J\bar J$, which are neutral under the center symmetry.  
Moreover, there must be a $\bZ_N$ symmetric relevant/marginal operator in between,
 \ie\label{range}
 N-1 \le c \le 9-N  \,,~~~N\le 4 .
 \fe 
See Table \ref{table:E8}.

For non-anomalous $\bZ_5$ at $c=4$, which is the value of the central charge extrapolated from \eqref{range} to $N=5$, the bound on the scalar primary operator is $\Delta_\text{scalar}^{Q=0} < 2.045$.  This bound is almost saturated by the marginal current bilinear operator in the $\mathfrak{su}(5)_1$ WZW model.
For $\bZ_6$ with the $k = 3$ anomaly, the range of $c$ is $3.98 < c < 4.85$, and the bound at $c = 5$ is $\Delta_\text{scalar}^{Q=0} < 2.022$, which is almost saturated by the $\mathfrak{su}(6)_1$ WZW model.  For $\bZ_6$ with other  anomalies and for $\bZ_7$ with any   anomaly, we could not establish any range of $c$ for which the bound drops to or below 2. 
This presents a natural stopping point for our numerical exploration.

Surprisingly, our numerical bootstrap bounds have an interesting connection to the $(\mathfrak{e}_8)_1$ vertex operator algebra (VOA). 
For each $ N \le 4$, the pair  $(\mathfrak{g}_1 , \mathfrak{h}_1)$ are both subVOAs of $(\mathfrak{e}_8)_1$  and they are the commutants of each other.  
Furthermore, the decomposition of $(\mathfrak{e}_8)_1$ into this pair of mutual commutants preserves precisely a $\bZ_N$ global symmetry with the anomaly described above.
The analogous decompositions 
for $N = 5, 6$ are into $\mathfrak{su}(5)_1 \times \mathfrak{su}(5)_1$ and $\mathfrak{su}(2)_1 \times \mathfrak{su}(3)_1 \times \mathfrak{su}(6)_1$, respectively.\footnote{We thank Theo Johnson-Freyd for an illuminating discussion on this point.} 
  Interestingly, our bootstrap bounds for non-anomalous $\bZ_5$ at $c = 4$ and that for $\bZ_6$ with the $k=3$ anomaly at $c = 5$ appear to be (almost) saturated by $\mathfrak{su}(5)_1$ and $\mathfrak{su}(6)_1$.\footnote{However, the bound for $\bZ_6$ with the $k=3$ anomaly at $c = 3$ is significantly above 2 and is not saturated by the $\mathfrak{su}(2)_1 \times \mathfrak{su}(3)_1$ WZW model.
}
We leave these curious observations and an analytic derivation of  \eqref{range} for future investigations.  
Some of these commutant pairs of $(\mathfrak{e}_8)_1$ have recently been discussed in \cite{Mukhi:2020sxt,Hegde:2021sdm}.

The ranges of $c$ from the other anomalies are wider than \eqref{range}. 
 Therefore, among the other constraints we have derived, we conclude that any CFT with a $\bZ_N$ symmetry in the range \eqref{range} must have a $\bZ_N$-symmetric relevant/marginal operator.

\begin{table}
\begin{align*}
\left.\begin{array}{|c|c|c|}\hline ~~\text{Symmetry}~~ & c &~~ \text{CFTs saturating the bounds} ~~\\\hline\hline ~~\text{None}~~ & 0\le c\le 8 &\big (~\text{trivial}~,~(\mathfrak{e}_8)_1 ~\big)\\\hline \bZ_2 & 1\le c\le 7 & \big(~ \mathfrak{su}(2)_1~,~(\mathfrak{e}_7)_1~\big ) \\\hline \bZ_3 &~~ 2\le c\le 6~~ & \big( ~\mathfrak{su}(3)_1~,~(\mathfrak{e}_6)_1~\big) \\\hline \bZ_4 & 3\le c \le 5 & \big( ~\mathfrak{su}(4)_1 = \mathfrak{so}(6)_1~,~\mathfrak{so}(10)_1~\big )
\\
\hline 
\end{array}\right.
\end{align*}
\caption{For $N \le 4$, we show that any CFT with a $\bZ_N$ global symmetry must have a $\bZ_N$-symmetric relevant/marginal operator in a range of $c$ that is numerically very close to that above. For each range of $c$, the bounds at the two ends are saturated by a pair of WZW models that can be embedded into $(\mathfrak{e}_8)_1$. 
 }
 \label{table:E8}
\end{table}

To check whether our bounds are correct, we will compute the $\bZ_N$ anomalies in a variety of (1+1)$d$ CFTs.  
An explicit formula for the 't Hooft anomaly can be derived from the following input data:\footnote{See \cite{2021arXiv210502909K} for a related discussion on the lattice. We thank Michael Levin for discussions on this point.}
\begin{itemize}
\item The Hilbert space $\cal H$ of local operators and their conformal weights $(h,\bar h)$.
\item The action of the $\mathbb{Z}_N$ global symmetry on the local operators.
\end{itemize}
First, from the input data one can readily compute $Z^\eta(\tau,\bar\tau)$, the torus partition function with a $\bZ_N$ symmetry operator  $\widehat\eta$ inserted at a fixed time. 
The $\bZ_N$ anomaly $k$, which is an element of $H^3(\bZ_N,{\rm U}(1))=\bZ_N$, can then be expressed as
\ie\label{mainformula}
e^{\frac{2\pi i k}{N}}  =  { Z^\eta( - {1\over \tau +N} , -{1\over\bar\tau+N}) \over Z^\eta( - {1\over \tau } , -{1\over\bar\tau})  }  \, .
\fe
Despite appearance, the right-hand side is actually independent of $\tau,\bar\tau$. 
It is important to stress that this formula for the anomaly holds true for general (1+1)$d$ bosonic CFT  with a unique vacuum, and is not restricted to rational or unitary CFT.  
We will apply this formula to  the $c=1$ compact boson, the WZW models, and the parafermion CFT in Section \ref{sec:example}.

This paper is organized as follows. 
 Section \ref{sec:symmetry} reviews the modern perspective on global symmetries and their anomalies in terms of the topological defects.  
In Section \ref{sec:example}, we compute the  't Hooft anomaly for various CFTs with $\bZ_N$ global symmetry. 
In  Section \ref{Sec:Bootstrap}, we set up the modular bootstrap system and present universal upper bounds on the lightest symmetry-preserving scalar operators.  
In Appendix~\ref{Sec:More}, we collect further details on the modular bootstrap system.
In Appendix \ref{App:Gap}, we present more general upper bounds on   charged local operators.

\section{Symmetries and Anomalies in (1+1)$d$}\label{sec:symmetry}

In this section, we start with a general characterization of global symmetries and their 't Hooft anomalies in (1+1)$d$ in terms of the topological defect lines.  
See \cite{Bhardwaj:2017xup,Tachikawa:2017gyf,Chang:2018iay,Lin:2019kpn,Thorngren:2019iar,Gaiotto:2020iye,Komargodski:2020mxz} and references therein for recent discussions on this topic.

\subsection{Topological Defect Lines}
\label{Sec:DefectHilbert}

The 0-form global symmetry in a general quantum field theory in $d$ spacetime dimensions is implemented by a codimension-one topological defect \cite{Kapustin:2014gua,Gaiotto:2014kfa}.  
When the global symmetry group is ${\rm U}(1)$, the topological defect is nothing but $\exp\left[ i \theta\int_\Sigma \star j\right]$, where $j_\mu(x)$ is the Noether current.  Here $\Sigma$ is a codimension-one closed manifold in spacetime (sometimes taken to be the whole space at a fixed time) where the topological defect has support.  
In this example, the topological nature of the defect follows from the conservation equation of the Noether current, $\partial^\mu j_\mu=0$, in flat spacetime with trivial background. 

For a discrete global symmetry where there is no local Noether current,  there  is still  a codimension-one topological defect  $\cL_g$ associated to every group element $g$.  
In the special case when the manifold of support $\Sigma$ is taken to be the whole space at a fixed time, $\cL_g$ is the charge operator of the symmetry. 
In Euclidean signature, we can encircle a local operator  $\phi(x)$ by  a topological defect $\cL_g$, and then  contract the defect without changing the correlation function.  This process produces another local operator $\phi'(x) = g\cdot \phi(x)$.\footnote{Strictly speaking, $\phi'(x) = \langle \cL_g \rangle_{\bR^d}\times  g\cdot \phi(x)$, where $\langle \cL_g \rangle_{\bR^2}$ is the vacuum expectation value of the topological defect in $\bR^d$, which might not always be +1.  For example, an anomalous $\bZ_2$ topological defect line in (1+1)$d$ has $\langle \cL \rangle_{\bR^2}=-1$.   See Section 2.4 of \cite{Chang:2018iay} and \cite{Cordova:2019wpi}.}

\begin{figure}
\centering
\begin{tikzpicture}[scale=.5]
\draw (0,0) ++ (180:2 and 1) arc (180:360:2 and 1);
\draw [dashed] (0,0) ++ (0:2 and 1) arc (0:180:2 and 1);
\draw (0,4) ellipse (2 and 1);
\draw (-2,0) -- (-2,4);
\draw (2,0) -- (2,4);
\draw [line width=0.5mm,->-=.52] (0,1) ++ (180:2 and 1) 
arc (180:360:2 and 1);
\node at (0,.55) {$\cal L~$} ;
\draw [line width=0.5mm,dashed] (0,1) ++ (0:2 and 1) arc (0:180:2 and 1);
\node at (0,-1.5) {$|\phi\rangle \in {\cal H}$};
\node at (4,1.5) {$\rightarrow$};
\begin{scope}[xshift = 8cm]
\draw (0,0) ++ (180:2 and 1) arc (180:360:2 and 1);
\draw [dashed] (0,0) ++ (0:2 and 1) arc (0:180:2 and 1);
\draw (0,4) ellipse (2 and 1);
\draw (-2,0) -- (-2,4);
\draw (2,0) -- (2,4);
\node at (0,-1.5) {$\widehat\cL |\phi\rangle \in {\cal H}$};
\end{scope}
\end{tikzpicture}
\caption{A topological defect line $\cal L$ wrapped around the spatial circle of a cylinder leads to an action $\widehat \cL$ on the Hilbert space $\cal H$. 
}
\label{fig:L}
\end{figure}
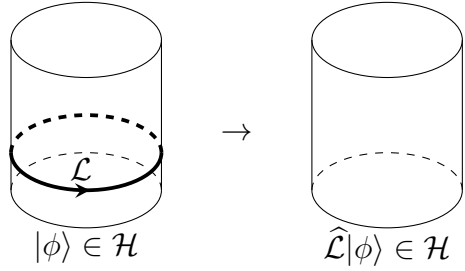

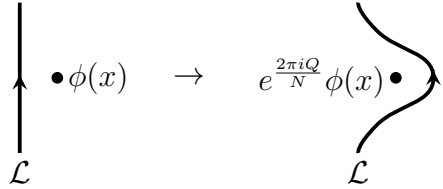
\begin{figure}
\centering
~~~~
\begin{tikzpicture}[scale=.5]
\draw [line width=0.5mm,->-=.52] (0,0) -- (0,4);
\node at (0,-.5) {$\cal L$};
\filldraw (1,2) circle (4pt) node[right=0mm] {$\phi(x)$};
\node at (4.5,2) {$\rightarrow$};
\begin{scope}[xshift=2cm]
\draw [line,->-=.5] (9,2.19) -- (9,2.21);
\draw [line width=0.5mm,smooth,tension=1] plot coordinates {(7,0) (7.5,.75) (9,2) (7.5,3.25) (7,4)};
\node at (7,-.5) {$\cal L$};
\filldraw (8,2) circle (4pt) node[left=0mm] {$e^{2\pi i Q \over N} \phi(x)$};
\end{scope}
\end{tikzpicture}
\caption{As a $\bZ_N$ TDL is swept past a local operator $\phi(x)$, the correlation function changes by a phase $e^{2\pi i Q \over N}$ where $Q=0,1,\cdots, N-1$ is the $\bZ_N$ charge of $\phi$.   }
\label{fig:0Z2}
\end{figure}

In (1+1)$d$, such codimension-one topological defects are lines. 
In this paper, we will focus on the topological defect line (TDL) $\cal L$ associated with an internal, unitary $\bZ_N$ symmetry in a bosonic (1+1)$d$ CFT.
The $\bZ_N$ TDL implements a  $\bZ_N$ action on the Hilbert space $\cal H$ quantized on a circle ${\rm S}^1$. 
This action can be realized on the cylinder ${\rm S}^1\times \bR$ by wrapping the $\bZ_N$ TDL around the compact (spatial) ${\rm S}^1$ direction, to act on a state $|\phi\rangle\in {\cal H}$ prepared at an earlier time (see Figure \ref{fig:L}).  We will denote this $\bZ_N$ unitary operator as
\begin{align}
\widehat\cL:~ {\cal H} \to {\cal H}\,. 
\end{align}
We can grade the Hilbert space $\cal H$ using the $\bZ_N$ charge $Q=0,1,\cdots, N-1$:
\ie
{\cal H} =\bigoplus_{Q=0}^{N-1} {\cal H}^{(Q)} \, .
\fe
Via the operator-state correspondence, the TDL also implements a $\bZ_N$ action on local operators: as a $\bZ_N$ TDL is swept past a local operator $\phi(x)$, the correlation function changes by a phase $e^{2\pi i Q \over N}$ (see Figure \ref{fig:0Z2}). 

The fusion of topological symmetry lines obeys the group multiplication law. 
In particular, as we bring $N$ parallel $\bZ_N$ loops  together, they fuse to a trivial line.  
Thus $\widehat\cL^N=1$.

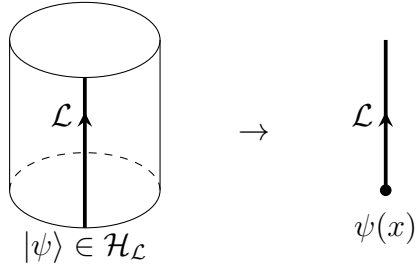
\begin{figure}
\centering
\begin{tikzpicture}[scale=.5]
\draw (0,0) ++ (180:2 and 1) arc (180:360:2 and 1);
\draw [dashed] (0,0) ++ (0:2 and 1) arc (0:180:2 and 1);
\draw (0,4) ellipse (2 and 1);
\draw (-2,0) -- (-2,4);
\draw (2,0) -- (2,4);
\draw [line width=0.5mm,->-=.77] (0,-1) -- (0,1) -- node[left] {$\cal L$} (0,3);
\node at (0,-1.5) {$|\psi\rangle \in {\cal H}_\cL$};
\node at (4.5,1.5) {$\rightarrow$};
\draw [line width=0.5mm,->-=.52] (8,0) -- node[left] {$\cal L$} (8,4);
\filldraw (8,0) circle (4pt);
\node at (8,-1) {$\psi(x)$};
\node at (11,-1) {};
\end{tikzpicture}
\caption{By quantizing the system on a spatial circle with a topological defect line $\cal L$ inserted at a point in space, we define  a defect Hilbert space ${\cal H}_\cL$.  Via the operator-state correspondence, the states in ${\cal H}_\cL$ are mapped to operators living at the end of $\cal L$.}
\label{fig:HL}
\end{figure}

Consider 
the theory on a cylinder ${\rm S}^1 \times \, \mathbb{R}$ with $\cal L$ running along the time $\mathbb{R}$ direction (see Figure \ref{fig:HL}).  
The $\cal L$ TDL intersects with the spatial ${\rm S}^1$, and therefore implements a twisted periodic boundary condition in the quantization.
This defines a {\it defect Hilbert space} denoted by ${\cal H}_\cL$.  
 Via the operator-state correspondence, a defect Hilbert space state $|\psi\rangle\in {\cal H}_\cL$ is mapped to an operator living at the end of the $\bZ_N$ TDL.

Since a TDL commutes with the stress tensor,  the states in the defect Hilbert space ${\cal H}_\cL$ can be organized into representations of the left and right Virasoro algebras. In particular,  the defect Hilbert space can be diagonalized to have definite conformal weights $(h,\bar h)$.

\subsection{Explicit Formula for $\bZ_N$ Anomalies}\label{sec:anomaly}

The 't Hooft anomaly of a global symmetry is characterized by certain  splitting-and-joining relations of their TDLs. 
More specifically, it is captured by the $F$-symbols of the fusion category formed by these TDLs. 
We refer the readers to  \cite{Bhardwaj:2017xup,Chang:2018iay,Lin:2019kpn,Thorngren:2019iar} for physicists' expositions on this subject. 

Here we focus on the case of a $\mathbb{Z}_N$ global symmetry in  (1+1)$d$ bosonic CFT.  
An explicit formula for the $\bZ_N$ anomaly can be derived from the input data of the conformal weights $(h,\bar h)$ and the $\bZ_N$ charges of all the  local operators in the Hilbert space $\cal H$.  
(The relation between modular invariance/covariance and anomalies has been discussed extensively in \cite{Freed:1987qk,Felder:1988sd,Gaiotto:2008jt,Sule:2013qla,Numasawa:2017crf}.)

The $\bZ_N$ anomaly is classified by the group cohomology $H^3(\mathbb{Z}_N,{\rm U}(1)) =\mathbb{Z}_N$ \cite{Freed:1987qk,Chen:2011bcp,Chen:2011pg}.  
Let $\eta$ be a generator of the $\mathbb{Z}_N$ global symmetry with 't Hooft anomaly $k$, which  is an integer modulo $N$.   
The cocycle of this group cohomology can be chosen to be (see, for example, \cite{Moore:1988qv})
\ie\label{Fsymbol}
\alpha(a,b,c) = \exp\left[{2\pi i k\over N^2}  a(b+c- \langle b+c\rangle ) \right]\,,
\fe
where $a,b,c=\{0,1,\cdots, N-1\}$ and $\langle a\rangle $ is the mod $N$ function to $\{0,1, \cdots , N-1\}$.

Consider the partition function $Z^\eta(\tau,\bar\tau)$ with the symmetry line $\eta$ wrapped around the spatial circle (a twist in the time direction) of a torus with complex modulus $\tau$.  
This partition function admits the following interpretation as a trace over the Hilbert space $\cal H$ of local operators with the $\bZ_N$ symmetry operator $\widehat\eta$ inserted:
\ie
\Z{1}{2}{}{}{} \quad Z^\eta(\tau, \bar\tau) = {\rm Tr}_{\cal H} \, \widehat\eta \, q^{h- {c\over24}} \bar q^{\bar h-{\bar c\over24}} \, ,
\fe
where $q = \exp(2\pi i \tau), \, \bar q = \exp(-2\pi i \bar\tau)$.  
Using the trace interpretation of the right-hand side, the partition function $Z^\eta(\tau,\bar\tau)$ can be computed by the  input data above. 
Under a modular S transform $\tau \to -1/\tau$, we obtain the torus partition function with an $\eta$ line running in the time direction,
\ie
\Z{1}{}{2}{}{} \quad Z_\eta(\tau, \bar\tau) = {\rm Tr}_{{\cal H}_\eta} \, q^{h-{c\over24}} \bar q^{\bar h-{\bar c\over24}}    =  Z^\eta(-1/\tau ,-1/\bar\tau)\, ,
\fe
where ${\cal H}_\eta$ is the defect Hilbert space introduced in Section~\ref{Sec:DefectHilbert}.

It was explained in \cite{Chang:2018iay} (see also \cite{Hung:2013cda}) that the spins $h-\bar h$ of the states in the defect Hilbert space ${\cal H}_\eta$ of such an anomalous $\bZ_N$ symmetry are constrained as
\ie
\label{SpinSelection}
{\cal H}_\eta:~~~~s\equiv h - \bar h \in {k \over N^2} + {\mathbb{Z} \over N} \, .
\fe
We will refer to this constraint as a spin selection rule.\footnote{This spin selection rule for a (possibly anomalous)  $\bZ_N$ symmetry in  (1+1)$d$ CFT is related to the spins of the anyons in the (2+1)$d$ $\bZ_N$ gauge theory (possibly with a Dijkgraaf-Witten twist \cite{Dijkgraaf:1989pz}).  See  \cite{Lin:2019kpn,Ji:2019eqo} for the case of $\bZ_2$ symmetry.} 

The spin selection rule \eqref{SpinSelection} implies that under the modular $T^N$ transformation, the partition function $Z_\eta$ of ${\cal H}_{\eta}$ is shifted by a phase determined by the anomaly:
\ie
Z_\eta(\tau+N,\bar \tau+N) =  e^{2\pi i \frac{k}{N}} Z_\eta(\tau,\bar\tau)\,,
\fe
Written in terms of $Z^\eta$, we find the following formula for the 't Hooft anomaly $k$:
\ie
e^{\frac{2\pi i k}{N}}  =  { Z^\eta( - {1\over \tau +N} , -{1\over\bar\tau+N}) \over Z^\eta( - {1\over \tau } , -{1\over\bar\tau})  }  \,.
\fe
In dividing the partition function, we have assumed that our CFT has a unique vacuum (\textit{i.e.}, there is a unique local operator with $h=\bar h=0$), so that $Z^\eta(\tau,\bar \tau)$ is not a zero function. 
We stress that  $Z^\eta(\tau,\bar\tau)$ is completely  and straightforwardly determined by the input data specified above.  
Note that despite appearance, the right-hand side is actually independent of $\tau,\bar\tau$.

Starting from a $\bZ_N$ symmetry and its anomaly, we now discuss the anomaly of its group extension and its subgroup.
A $\bZ_N$ symmetry can be viewed as an unfaithful $\bZ_{pN}$ symmetry,  for some positive integer $p$.  
Mathematically, this means that the $\bZ_N$ symmetry is {\it extended} to $\bZ_{pN}$, where $\eta$ is now regarded as an order $pN$ element, $\eta^{pN}=1$.  
By viewing the spin selection rule \eqref{SpinSelection} as that for a $\bZ_{pN}$ defect Hilbert space, we find that the anomaly of this $\bZ_{pN}$ symmetry generated by $\eta$ is $p^2 k $ mod $pN$. 
In particular, by choosing $p=N/\text{gcd}(k,N)$, the $\bZ_N$ anomaly is trivialized when extended to $\bZ_{pN}$.
See \cite{Wang:2017loc,Tachikawa:2017gyf} for a more general discussion on how anomalies can be trivialized via extension.

Next, consider the $\bZ_{N/\text{gcd}(r,N)}$ subgroup of $\bZ_N$ generated by $\eta^r$ for some positive integer $r$.  
 The spin selection rule for the defect Hilbert space of $\eta^r$ is 
\ie
\label{SpinSelectionAut}
{\cal H}_{\eta^r}:~~~~s\in{r^2 k \over N^2} + {\mathbb{Z} \over N/\text{gcd}(r, N)}  \, .
\fe
Hence the anomaly of the $\bZ_{N/\text{gcd}(r,N)}$ subgroup generated by $\eta^r$ is $\left( {r\over \text{gcd}(r,N)}\right)^2 k$~ mod~ $N\over \text{gcd}(r,N)$.

The torus partition functions $Z^{\eta^r}$ and $Z_{\eta^r}$ form a system closed under the modular S transform. 
The consistency of the modular S transform can then be exploited to study and constrain different charge sectors.
This is the subject of Section~\ref{Sec:Bootstrap}.

\section{Examples of $\bZ_N$ Symmetries and Anomalies}\label{sec:example}

In this section we review and compute the $\bZ_N$ anomalies in various different CFTs.

\subsection{Anomalies in Diagonal Rational CFTs}

In rational CFT with diagonal modular invariance, there is a natural class of TDLs known as the \textit{Verlinde lines} \cite{Verlinde:1988sn,Petkova:2000ip}.  
They have the distinguished property that they commute  not only with the Virasoro algebra, but also  with the entire left and right chiral algebras.  
Modular covariance constrains the Verlinde lines to be in one-to-one correspondence with the primaries of the chiral algebra.  

Let us review the action of the Verlinde lines on the Hilbert space. 
Let $|\phi_i\rangle$ be the chiral algebra primary of a rational CFT, with the index $i$ labels different primaries. 
The Verlinde line $\cL_k$ associated with the primary $\phi_k$ acts on the primary   $|\phi_i\rangle$ as
\ie
\widehat\cL_k |\phi_i \rangle = {S_{ki}\over S_{0i} }|\phi_i\rangle\,,
\fe
where $S_{ki}$ is the modular S-matrix and $0$ stands for the identity operator.  
In particular, the Verlinde lines in a diagonal RCFT commutes with parity, which exchanges $h$ with $\bar h$.

The Verlinde lines, such as the Kramers-Wannier duality line in the Ising CFT \cite{Frohlich:2004ef,PhysRev.60.252}, are generally non-invertible.  
We will refer to the invertible ones as the \textit{Verlinde symmetry lines}.  
Simple examples of symmetries that are realized by Verlinde lines are the $\bZ_2$ symmetry in the Ising CFT, and the  center symmetry in the  WZW model that commutes with the left and the right current algebras.

Using the spin selection rule, we will now show that certain 't Hooft anomalies cannot be realized by Verlinde symmetry lines. 
Since the Verlinde symmetry lines commute with parity, their defect Hilbert spaces should be invariant under  flipping  the sign of the spin $s = h-\bar h \to -s$.
We conclude that an anomaly can be realized by the Verlinde symmetry lines only if the spin selection rules of its defect Hilbert spaces are invariant under $s \to -s$.

Consider the special  case of a $\bZ_N$ global symmetry generated by $\eta$  with anomaly $k$.
Let $\mathcal{S}$ be the set of spins in the defect Hilbert space ${\cal H}_\eta$. 
Every element $s$ of $\mathcal{S}$ is constrained by the spin-selection rule  to satisfy \eqref{SpinSelection}. 
It follows that a necessary condition for the set $\mathcal{S}$ to be invariant under $s\to -s$ is 
\ie\label{2kN}
2k\in N\mathbb{Z}
\, .
\fe 
This is equivalent to
\ie\label{UBTC}
&k=  0~~~\text{mod}~N\,,~~~~\text{for odd}~N\,,\\
&k=  0,{N\over 2}~~~\text{mod}~N\,,~~~~\text{for even}~N\,.
\fe
We conclude that a $\bZ_N$ symmetry with  anomaly $k$ can be realized by Verlinde symmetry lines only if $2k\in N\mathbb{Z}$.  
For small $N$, we tabulate the possible $\bZ_N$ anomalies that can be realized by the Verlinde symmetry lines in certain diagonal RCFTs in Table~\ref{Tab:Examples}.

In any diagonal RCFT, the set of Verlinde lines can be lifted into a UMTC.\footnote{The relative 't Hooft anomaly in RCFT has recently been analyzed from this perspective in \cite{Cheng:2020rpl}.}   
 In particular, the Verlinde symmetry lines, which are the subset of invertible Verlinde lines, can be lifted to a unitary braided tensor category (UBTC). 
 The constraints \eqref{2kN} imply that not every 't Hooft anomaly, which is  encoded in the $F$-symbols \eqref{Fsymbol} of the TDLs, is compatible with an UBTC.  
It is indeed known that only  those $F$-symbols obeying \eqref{UBTC} can be lifted to a UBTC \cite{Moore:1988qv}.   
Above we provide a derivation of this well-known fact from a different perspective.

 For instance, the anomalous $\bZ_3$ symmetry cannot be lifted to a UBTC, and therefore cannot be realized by the Verlinde lines in a diagonal RCFT.  
 As another example, the $\bZ_4$ symmetries with $k=1,3$ anomalies in Section \ref{sec:Tduality} also cannot be lifted to a UBTC.

\begin{table}
\begin{equation*}
\begin{aligned}
\left.\begin{array}{|c|c|c|}\hline G & \text{Anomaly}~k & \text{RCFTs} \\
\hline ~~\bZ_2~~ & 0 & \text{Ising}, ~\mathfrak{su}(2)_{2n}  \\
\hline & 1&\mathfrak{su}(2)_{2n+1}\\
\hline \bZ_3 & 0 & \text{$\bZ_3$ parafermion},~\mathfrak{su}(3)_n \\
\hline \bZ_4 & 0 & \text{$\bZ_4$ parafermion},~\mathfrak{su}(4)_{2n} \\
\hline  & 2 & \mathfrak{su}(4)_{2n+1}, \, \mathfrak{so}(4n+2)_1 \\
\hline \bZ_5 & 0&\text{$\bZ_5$ parafermion},~\mathfrak{su}(5)_n  \\
\hline \end{array}\right.
\end{aligned}
\end{equation*}
\caption{$\bZ_N$ anomalies that can be realized as Verlinde symmetry lines in diagonal RCFTs for $N=2,\cdots,5$.  On the third column we tabulate  some diagonal RCFTs  that realize these anomalies. Here $n$ is any positive integer. }
\label{Tab:Examples}
\end{table}

\subsection{$c=1$ Compact Boson}

The $c=1$ compact boson at radius $R$ is described by a target space field $X(z, \bar z)$ with periodicity $X(z, \bar z) \sim X(z, \bar z) + 2\pi R$ and the OPE\footnote{Our convention for the radius $R$ is such that $R=1$ is the self-dual point with $\mathfrak{su}(2)_1$ current algebra.}
\ie
X(z, \bar z) X(0, 0) \sim - {1\over2} \log |z|^2 \, .
\fe
The left and right current algebras are generated by the holomorphic and anti-holomorphic currents $i \partial X(z)$ and $i \bar\partial X(\bar z)$, which have the OPE
\ie
i \partial X(z) \ i \partial X(0) \sim {1 \over 2z} \, , \quad i \bar\partial X(\bar z) \ i \bar\partial X(0) \sim {1 \over 2\bar z} \, .
\fe
The current algebra primaries are exponential operators labeled by a momentum number $n \in \bZ$ and a winding number $w \in \bZ$, 
\ie
{\cal O}_{n,w} (z, \bar z) = {:} \exp\left[ i ({n \over R}+wR) X_L(z) + i ({n \over R}-wR) X_R(\bar z) \right] {:}
\fe
with conformal weights
\ie
h = \frac 14 ({n \over R}+wR)^2 \, ,~~~\bar h = \frac 14 ({n \over R}-wR)^2 \,.
\fe
Here $:~ :$ stands for normal-ordering of the operator. 
The scaling dimension and the spin are
\ie
\Delta =h+\bar h= \frac 12 ({n^2 \over R^2}+w^2R^2) \, ,~~~s= h-\bar h= nw \, .
\fe

The OPE between the (anti-)holomorphic currents and the primaries is
\ie
i \partial X(z) {\cal O}_{n,w} (0, 0) \sim {{n \over R} + wR \over 2z} {\cal O}_{n,w} (0,0) \, , \quad i \bar\partial X(\bar z) {\cal O}_{n,w} (0,0) \sim {{n \over R} - wR \over 2\bar z} {\cal O}_{n,w} (0,0) \, .
\fe
The Ward identity then implies that at an irrational $R^2$, the exponential operators have mutually irrational charges under each current, {\it i.e.} $i \partial X(z)$ and $i \bar\partial X(\bar z)$ each generates a non-compact $\bR$ instead of a compact ${\rm U}(1)$ global symmetry.  Nonetheless, the symmetry group is a compact ${\rm U}(1) \times {\rm U}(1)$ torus.  The compact cycles are generated by the momentum and winding currents that have holomorphic and anti-holomorphic components
\ie
\text{Momentum } {\rm U}(1)_{n}: & \quad J_n(z) = i R \partial X(z), \quad \bar J_n(z) = i R \bar\partial X(\bar z) \, ,
\\
\text{Winding } {\rm U}(1)_{w}: & \quad J_w(z) = {i \over R} \partial X(z), \quad \bar J_w(z) = - {i \over R} \bar\partial X(\bar z) \, ,
\fe
under which the exponential operators ${\cal O}_{n,w} (z, \bar z)$ have integer charges $n$ and $m$, respectively.

The ${\rm U}(1)_{n}$ and ${\rm U}(1)_{w}$ symmetries are each non-anomalous by themselves, but they have a mixed anomaly.  
For a pair of coprime integers $p, q$, consider the compact ${\rm U}(1)_{p,q}$ subgroup generated by a Noether current with holomorphic and anti-holomorphic components
\ie
& J_{p,q}(z) = p J_{n}(z) + q J_{w}(z) = \left( pR+{q \over R} \right) i \partial X(z) \, , 
\\
& \bar J_{p,q}(\bar z) = p \bar J_{n}(\bar z) + q \bar J_{w}(\bar z) = \left( pR - {q \over R} \right) i \bar\partial X(\bar z) \, .
\fe
Its anomaly is given by $k_{{\rm U}(1)} = {1\over2} \left[ {(pR+{q \over R})^2 \over 2} - {(pR-{q \over R})^2 \over 2} \right] = pq$.\footnote{We follow the convention that the ${\rm U}(1)$ anomaly $k_{{\rm U}(1)}$ in a bosonic quantum field theory is an integer, and is related to the two point function of the Noether currents by
\ie
\langle J(z) J(0) \rangle = {k \over z^2} \, , \quad \langle \bar J(\bar z) \bar J(0) \rangle = {\bar k \over \bar z^2}, \quad k_{{\rm U}(1)} = {k - \bar k \over 2} \in \bZ \, .
\fe
}
Generally, given a ${\rm U}(1)$ symmetry with $k_{{\rm U}(1)}$ anomaly, its $\bZ_N$ subgroup has $k_N \equiv k_{{\rm U}(1)} \mod N$ anomaly.  To summarize, the ${\rm U}(1)$ anomaly and the anomaly of its $\bZ_N$ subgroup are given by
\ie
k_{{\rm U}(1)} = pq \, , \quad k_N \equiv pq \mod N \, .
\fe
Therefore, the $c=1$ compact boson at any definite radius $R$ realizes all possible ${\rm U}(1)$ and cyclic anomalies.\footnote{The exponential operators ${\cal O}_{n,w}(z, \bar z)$ have charge $Q_{p,q} = p n + q w$ under ${\rm U}(1)_{p,q}$.  With respect to $\bZ_N < {\rm U}(1)_{p,q}$, the uncharged sector consists of the exponential operators with $Q_{p,q} = p n + q w \equiv 0 \mod N$, and the uncharged scalar gap in the spectrum of exponential operators (hence excluding $J \bar J$ which has scaling dimension 2 and all other current algebra descendants) is given by
\ie
\label{c=1Gap}
\Delta_\text{scalar,exp}^{Q=0} = {1\over2} \, \min_R \left[ \, {N^2 \over \text{gcd}(p, N)^2 R^2}, \, { \text{gcd}(q, N)^2 R^2 \over N^2 } \, \right] \, .
\fe
Whenever $\Delta^{Q=0}_\text{scalar,exp}$ is greater than 2, there is a range of $R$ such that the $c=1$ compact boson is robust against perturbation preserving this $\bZ_N$ symmetry with anomaly $k_N = pq$ mod $N$.
}

\subsection{Center Symmetries in WZW Models}

We now consider the center symmetries in the $\mathfrak{su}(N)_1$, $\mathfrak{so}(4N+2)_1$, and $(\mathfrak{e}_6)_1$ WZW models.  
The centers of the other WZW models with simple Lie algebras are either trivial or products of $\bZ_2$, which have already been considered in \cite{Lin:2019kpn}.  
We focus on WZW models at level 1 because some of them turn out to saturate our bounds. 
The anomaly of the center symmetry for general WZW models have also been considered in   \cite{Numasawa:2017crf}.

For the above WZW models at level 1, the center symmetry lines constitute all the Verlinde lines. Therefore, the fusion category of the $\bZ_N$ center symmetry can be lifted to a UMTC. 
It follows that the anomaly $k$ of the $\bZ_N$ center symmetry, which is encoded in the $F$-symbols, is determined to be $k=0$ if $N$ is odd, and $k=N/2$ if $N$ is even \cite{Moore:1988qv}. 
Below we confirm this  by a (1+1)-dimensional computation using the algorithm in Section \ref{sec:anomaly}.

\subsubsection{$\mathfrak{su}(N)_1$}

The $\mathfrak{su}(N)_1$ WZW model has central charge $c = N-1$ and center symmetry $\bZ_N$. It contains $N$ current algebra primaries with $\bZ_N$ charges $Q = 0, 1, \dotsc, N-1$ and weights
\ie
h_Q = \bar h_Q = {Q(N-Q) \over 2N} \, .
\fe
The modular $S$ matrix is
\ie
S_{QQ'} = {\omega^{QQ'} \over \sqrt{N}} \, , \quad \omega = e^{2\pi i \over N} \, .
\fe
In the spectrum of {\it Virasoro} primaries, the gap in each charge $Q$ sector is
\ie
\Delta^Q = 
\begin{cases}
2 & Q = 0 \, ,
\\
{Q(N-Q) \over N} & Q \neq 0 \, .
\end{cases}
\fe
Here the lightest $\bZ_N$-symmetric Virasoro primary with $\Delta^{Q=0}=2$ is the current bilinear operator.

Denote the $\mathfrak{su}(N)_1$ characters (without flavor fugacity) by $\chi_Q(\tau)$.  
For later convenience, we  define the subscript modulo $N$, {\it i.e.} $\chi_{Q+N}(\tau) = \chi_Q(\tau)$.
The torus partition function with no twist is
\ie
Z(\tau, \bar\tau) = \sum_{Q = 0}^{ N-1} |\chi_Q(\tau)|^2 \, . 
\fe

Let $\eta$ be a generator of $\bZ_N$.  
The torus partition function with a $\bZ_N$ twist in the time direction is
\ie
Z^{\eta}(\tau, \bar\tau) = \sum_{Q = 0}^{N-1} \omega^{Q} |\chi_Q(\tau)|^2 \, .
\fe
Performing a modular S transform on the latter gives the torus partition function with a $\bZ_N$ twist in the spatial direction
\ie
Z_{\eta}(\tau, \bar\tau) = \sum_{Q = 0}^{ N-1} \chi_Q(\tau) \overline{\chi_{Q+1}(\tau)} \, .
\fe
Therefore, the spins of the operators in the defect Hilbert space  ${\cal H}_\eta$ are
\ie
s=  {Q \over N} - {N-1  \over2 N} \, ,
\fe
with $0 \le Q \le N-1$.
Comparing these spins with the spin selection rule \eqref{SpinSelection}, we have found the anomaly of the $\bZ_N$ center symmetry in the $\mathfrak{su}(N)_1$ WZW model:
\ie
k= 
\begin{cases}
0~~~~N~\text{odd}\\
{N\over2}~~~N~\text{even}\,.
\end{cases}
\fe

\subsubsection{$\mathfrak{so}(4N+2)_1$}
\label{Sec:SO4N+2}

The $\mathfrak{so}(4N+2)_1$ WZW model has central charge $c = 2N+1$ and a $\bZ_4$ center symmetry $\bZ_4$.\footnote{The $\mathfrak{so}(4N)_1$ WZW model, by contrast, has a $\bZ_2\times \bZ_2$ center symmetry.
}
It is the bosonization of $4N+2$ free Majorana fermions. 
There are four current algebra primaries with $\bZ_4$ charges $Q = 0, 1, 2, 3$ and weights
\ie
h_0 = \bar h_0 = 0 \, , \quad h_1 = \bar h_1 = h_3 = \bar h_3 = {2N+1\over8} \, , \quad h_2 = \bar h_2 = {1\over2} \, .
\fe
The modular $S$ matrix is 
\ie
S = \frac12 \begin{pmatrix}
 1 & 1 & 1 & 1 \\
 1 & i & -1 & -i \\
 1 & -1 & 1 & -1 \\
 1 & -i & -1 & i
\end{pmatrix} \quad \text{for odd } N \, ,
\qquad
S = \frac12 \begin{pmatrix}
 1 & 1 & 1 & 1 \\
 1 & -i & -1 & i \\
 1 & -1 & 1 & -1 \\
 1 & i & -1 & -i
\end{pmatrix} \quad \text{for even } N \, .
\fe
In the spectrum of {\it Virasoro} primaries, the gaps in each charge $Q$ sector are
\ie
\Delta^Q = 
\begin{cases}
2 & Q = 0 \, ,
\\
{2N+1 \over 4}
& Q = 1, \, 3 \, ,
\\
1 & Q = 2 \, .
\end{cases}
\fe

The torus partition function with no twist is
\ie
Z(\tau, \bar\tau) = \sum_{Q = 0, 1, 2, 3} |\chi_Q(\tau)|^2 \, ,
\fe
The torus partition function with a $\bZ_4$ twist in the time direction is ($\eta$ is a generator of $\bZ_4$)
\ie
Z^\eta(\tau, \bar\tau) = |\chi_0(\tau)|^2 - |\chi_2(\tau)|^2 \, .
\fe
Under a modular S transformation, we find 
\ie
Z_\eta(\tau, \bar\tau) 
&= [ \, \chi_0(\tau) + \chi_1(\tau) \, ] \, \overline{\chi_2(\tau)} + \chi_2(\tau) \, [ \, \overline{ \chi_0(\tau) + \chi_1(\tau) } \, ] \, ,
\fe
which has spin content
\ie
\label{SOSpins}
h - \bar h = \left\{ \pm {1\over2}, \, \pm {2N-3 \over 8} \right\}.
\fe
Comparing with \eqref{SpinSelection}, we see that the $\bZ_4$ center has anomaly
\ie
k = 2 \, .
\fe

\subsubsection{$(\mathfrak{e}_6)_1$}

The $(\mathfrak{e}_6)_1$ WZW model has central charge $c = 6$ and a $\bZ_3$ center symmetry.  It has three current algebra primaries with $\bZ_3$ charges $Q = 0, 1, 2$ and weights
\ie
h_0 = \bar h_0 = 0 \, , \quad h_1 = \bar h_1 = h_2 = \bar h_2 = {2\over3} \, .
\fe
In the spectrum of {\it Virasoro} primaries, the gaps in each charge sector are
\ie
\Delta^Q = 
\begin{cases}
2 & Q = 0 \, ,
\\
{4\over3} & Q = 1, \, 2 \, .
\end{cases}
\fe
The $S$ matrix is the complex conjugate of $\mathfrak{su}(3)_1$,
\ie
S = \frac{1}{\sqrt3} \begin{pmatrix}
1 & 1 & 1
\\
1 & \omega^2 & \omega
\\
1 & \omega & \omega^2
\end{pmatrix} \, , \quad \omega = e^{2\pi i \over 3} \, .
\fe
Hence, the determination of the spin content proceeds in a similar fashion to the case of $\mathfrak{su}(3)_1$:
\ie
Z^{\eta}(\tau, \bar\tau) &= \sum_{Q = 0}^2 \omega^{Q} |\chi_Q(\tau)|^2 \, ,
\\
Z_{\eta}(\tau, \bar\tau) &= \sum_{Q = 0}^2 \chi_Q(\tau) \overline{\chi_{Q-1}(\tau)} \, , \quad s = \frac13, \, \frac23, \, 0 \, .
\fe
Comparing with \eqref{SpinSelection}, we conclude that the $\bZ_3$ center of $(\mathfrak{e}_6)_1$ is non-anomalous.

\subsection{T-duality $\bZ_4$ in $\mathfrak{su}(2)_1$}\label{sec:Tduality}

In this subsection we discuss examples of anomalous $\bZ_4$ symmetries in the $\mathfrak{su}(2)_1$ WZW model.  
These $\bZ_4$ symmetries have the $k=1$ or $k=3$ anomalies, which according to \eqref{2kN},  cannot be realized as Verlinde symmetry lines, \textit{i.e.}, they do not commute with the  maximally extended chiral algebra.

At a generic radius of the $c=1$ compact boson, T-duality is not a global symmetry, but a map of one description of the theory with radius $R$ to a different description of the same theory with radius $1/R$.     
See \cite{Fuchs:2007tx} for discussions on T-duality and  topological interfaces in the compact boson theory.

However, at the self-dual point $R=1$, {\it i.e.} the $\mathfrak{su}(2)_1$ WZW model, the T-duality becomes a global symmetry of the theory.  
The right T-duality is the chiral $\pi$ rotation of the right $\mathfrak{su}(2)_1$ current algebra  and is an order 4 action \cite{Harvey:2017rko}.
It is not a Verlinde symmetry because it acts nontrivially on the right $\mathfrak{su}(2)_1$ algebra.

Similarly, there is another order 4 action corresponding to the chiral $\pi$ rotation of the left $\mathfrak{su}(2)_1$ current algebra, which we call the left T-duality. 
The square of either $\bZ_4$ is the anomalous $\bZ_2$ center symmetry.

Let us describe the right
T-duality $\bZ_4$ in more detail, and compute its anomaly. 
The $\mathfrak{u}(1)$ current algebra primaries are labeled by a momentum number and a winding number,
\ie
&{\cal O}_{n,w} (z,\bar z) = \exp\left[   i (n+w)  X_L(z)  + i (n-w)X_R(\bar z)\right]\,,\\
&h = \frac 14 (n+w)^2 \,,~~~\bar h = \frac 14 (n-w)^2\,.
\fe
The left and the right $\mathfrak{su}(2)_1$ current algebra generators are
\ie
J_L^3(z)  =i  \partial X_L(z)\,,~~~~J_L^\pm (z) = {\cal O}_{\pm1, \pm1}= e^{\pm 2iX_L(z) }\,,\\
 J_R^3(\bar z)  =i \partial X_R(\bar z)\,,~~~~ J_R^\pm (\bar z) = {\cal O}_{\pm1, \mp1}= e^{\pm 2iX_R(\bar z) }\,.
\fe
There are two current algebra primaries, the vacuum primary $|h=0,\bar h=0\rangle$ and the spin-$\frac12$ primary $|h=\frac14,\bar h=\frac 14\rangle$. 
Those ${\cal O}_{n,w}$ with $n=w$ mod 2 belong to the vacuum module, while those with $n=w+1$ mod 2 belong to the spin-$\frac12$ module. 

The right T-duality $\bZ_4$ acts on $X_L$, $X_R$, and ${\cal O}_{n,w}$ as \cite{Harvey:2017rko}
\ie
&X_L(z) \to X_L(z)\,,~~~~X_R(\bar z) \to -X_R(\bar z)\,,\\
&{\cal O}_{n,w}  \to e^{ {i\pi\over2}  (n+w)^2 } {\cal O}_{w,n}\,. 
\fe
Note that this $\bZ_4$ does not commute with the right-moving $\mathfrak{u}(1)$ current algebra generated by $\bar\partial X_R(\bar z)$. 
The square of the generator of this $\bZ_4$ acts on ${\cal O}_{n,w}$ as 
\ie
\bZ_2:~{\cal O}_{n,w} \to (-1)^{n+w} {\cal O}_{n,w}\,,
\fe
which is the anomalous $\bZ_2$ center symmetry of the $\mathfrak{su}(2)_1$ WZW model (see, for example, \cite{Lin:2019kpn}).

The torus partition function of the $\mathfrak{su}(2)_1$ WZW model without any defect is
\ie
Z  (\tau,\bar \tau ) = {\rm Tr}_{\cal H} \, [q^{h -{ 1\over24} }\bar q^{\bar h -{1\over 24}} ]
=
\sum_{n, w \in \bZ} {q^{\frac14(n+w)^2 - {1\over 24}} \bar q^{\frac14(n-w)^2 - {1\over 24}} \over \prod_{m=1}^\infty (1-q^m)(1-\bar q^m)}
= \left| {\theta_3(2\tau) \over \eta(\tau) } \right|^2+ \left| {\theta_2(2\tau) \over \eta(\tau) } \right|^2\,.
\fe

We now  compute the partition function $Z^\eta$ with a right T-duality $\bZ_4$ line $\eta$ extended along the spatial direction.  
Since $\bZ_4$ swaps ${\cal O}_{n,w}$ with ${\cal O}_{w,n}$, only those terms with 
$n=w$ contribute. 
Furthermore, the right-moving current algebra oscillators contribute with a minus sign to the partition function.
Hence we have
\ie
Z^\eta(\tau,\bar \tau ) &=  \Tr_{\cal H}[\, \widehat\eta\, q^{h -{ 1\over24} }\bar q^{\bar h -{1\over 24}} ] = \sum_{n\in \bZ} {  q^{n^2 - {1\over 24} } \bar q^{-{1\over 24}   } \over \prod_{m=1}^\infty (1-q^m)(1+\bar q^m)}
= {\theta_3(2\tau) \over \eta(\tau) }  \sqrt{ 2\overline{\eta(\tau)} \over \overline{\theta_2(\tau)}} \,.
\fe
Its modular S transform is the partition function of the $\bZ_4$ defect Hilbert space:
\ie
Z_\eta(\tau,\bar \tau)& = \Tr_{{\cal H}_\eta}  [\, q^{h -{1\over 24} } \bar q^{\bar h -{1\over 24}}\, ] = \left( {\theta_3(2\tau) +\theta_2(2\tau) \over \eta(\tau) }\right) \sqrt{ \overline{\eta(\tau)}\over \overline{\theta_4(\tau)}}\,,\\
&= {1\over |\eta(\tau)|^2}  \left( \,
\bar q^{1\over 16}  + 2q^{1\over 4} \bar q^{1\over 16}  +\bar q^{9\over 16}  +2q^{1\over 4} \bar q^{9\over 16}
+2 q \bar q^{1\over 16}+\cdots
\,\right) \, .
\fe
From the spins 
of these operators, we conclude that the T-duality $\bZ_4$ has the $k=3$ anomaly.  
Similarly, the left T-duality $\bZ_4$, which acts by the chiral $\pi$ rotation of the left $\mathfrak{su}(2)_1$ current algebra, has the $k=1$ anomaly.

\subsection{$\bZ_N$ Parafermion CFT}

The $\bZ_N$ parafermion CFT, also known as the ${\mathfrak{su}(2)_N \over \mathfrak{u}(1)}$ coset CFT, 
is a $c = {2(N-1) \over N+2}$ diagonal RCFT \cite{Fateev:1985mm}.  It consists of $N(N+1) \over 2$ primaries labeled by two integers $(\ell, m)$ in the range 
\ie
0 \le \ell \le N, \quad -\ell+2 \le m \le \ell, \quad \ell - m \in 2\bZ \, ,
\fe
with conformal weights
\ie
h = \bar h = {\ell(\ell+2) \over 4(N+2)} - {m^2 \over 4N} \, .
\fe
There is a non-anomalous $\bZ_N$ symmetry that commutes with the parafermion algebra, and the primary labeled by $(\ell, m)$ has $\bZ_N$ charge $m$ mod $N$.  The uncharged sector consists of the identity ($\ell = m = N$) and all the primaries with $m = 0$.  The lightest nontrivial (scalar) operator in the uncharged sector is the $\ell = 2, \, m = 0$ primary, giving
\ie
\Delta_\text{scalar}^{Q=0} = {4 \over N+2} \, .
\fe
The first few parafermion CFTs with $N = 2, 3, 4$ are the familiar $c={1\over2}$ Ising CFT, the $c={4\over5}$ three-state Potts CFT, and the  $c=1$ free boson orbifold $S^1/\bZ_2$ at radius $R = \sqrt3$.

Let us explicitly verify that the $\bZ_N$ symmetry is non-anomalous by analyzing the spin contents in $Z_\eta$. 
We start with
\ie
Z^\eta(\tau, \bar\tau) = \sum_{\ell, m} e^{2 \pi i m \over N} \chi_{\ell, m}(\tau) \overline{\chi_{\ell, -m}(\tau)} \, .
\fe
Using the modular $S$ matrix \cite{Gepner:1986hr}
\ie
{S_{\ell, m}}^{\ell', m'} = {2 \over \sqrt{N(N+2)}} \sin {\pi (\ell+1) (\ell'+1) \over N+2} e^{ \pi i m m' \over N }
\fe
we obtain the partition function in the defect Hilbert space
\ie
Z_\eta(\tau, \bar\tau) &= \sum_{\ell, m, \bar\ell, \bar m} \left[ \sum_{\ell', m'} {S_{\ell', m'}}^{\ell, m} e^{2 \pi i m' \over N} {S_{\ell', -m'}^*}^{\bar\ell, \bar m} \right] \chi_{\ell, m}(\tau) \overline{\chi_{\bar\ell, \bar m}(\tau)}
\\
&= \sum_{\ell, m, \bar\ell, \bar m} n_{\ell, m; \bar\ell, \bar m} \chi_{\ell, m}(\tau) \overline{\chi_{\bar\ell, \bar m}(\tau)} \, ,
\fe
where
\ie
n_{\ell, m; \bar\ell, \bar m} 
= {4 \over N(N+2)} \sum_{\ell', m'} \sin {\pi (\ell+1) (\ell'+1) \over N+2} \sin {\pi (\bar\ell+1) (\ell'+1) \over N+2} e^{\pi i m'(m - \bar m + 2) \over N}
\fe
Using the fact that the summand is invariant under the simultaneous shift of $\ell' \to N - \ell', \, m' \to m' + N$,
we can rewrite the sum as ($m' = 2p'$)
\ie
\sum_{\ell', m'} \quad &\to \quad {1\over2} \sum_{\ell' = 0}^N \left( \sum_{p' = -{\ell'\over2}+1}^{\ell'\over2} + \sum_{p' = -{N-\ell'\over2}+1+{N\over2}}^{{N-\ell'\over2}+{N\over2}} \right)
\quad \to \quad {1\over2} \sum_{\ell' = 0}^N \sum_{p' = -{\ell'\over2}+1}^{N-{\ell'\over2}} \, ,
\fe
and furthermore, 
\ie
\sum_{p' = -{\ell'\over2}+1}^{N-{\ell'\over2}} e^{2 \pi i p'(m - \bar m + 2) \over N} 
&= 
\begin{cases}
N & \bar m = m + 2 \mod 2N \, ,
\\
- (-)^{N-\ell'} N & \bar m = m + 2 + N \mod 2N \, ,
\\
0 & \text{otherwise} \, .
\end{cases}
\fe
With this, one can easily carry out the sum over $\ell'$ to arrive at
\ie
n_{\ell, m; \bar\ell, \bar m}
&= 
\begin{cases}
1 & \bar\ell = \ell \text{~~and~~} \bar m \equiv m + 2 \mod 2N \, ,
\\
1 & \bar\ell = N - \ell \text{~~and~~} \bar m \equiv m + 2 + N \mod 2N \, ,
\\
0 & \text{otherwise} \, .
\end{cases}
\fe
It follows that the spins in $Z_\eta$ are always  $ {m+1 \over N}  +\bZ$, in agreement with the spin selection rule for a non-anomalous $\bZ_N$ symmetry.

\section{Modular Constraints on Symmetry-Preserving Scalar Operators}
\label{Sec:Bootstrap}

\subsection{Modular Bootstrap}

To apply the bootstrap algorithm, we need to identify a set of observables $\bf Z$ that (a) has an expansion on a certain basis of functions with non-negative coefficients, and (b) obeys a certain crossing equation. 
For our application, the observables $\bf Z$ will be  linear combinations of the torus partition functions twisted by different symmetry lines.  
We will pick a basis for these partition functions so  that they have non-negative expansions in the Virasoro characters. 
The crossing equation will come from the modular S transformation. 
Below we discuss in detail these partition functions and the crossing equation they satisfy.

Focusing on $c>1$, the Virasoro character $\chi_h(\tau)$ is non-degenerate when $h > 0$, and degenerate when $h = 0$,
\ie
\label{Character}
\chi_h(\tau) =
\begin{cases}
\displaystyle
{q^{h-\frac{c-1}{24}} \over \eta(\tau)} & h > 0 \, ,
\\
\\
\displaystyle
{q^{h-\frac{c-1}{24}} \over \eta(\tau)} (1-q) & h = 0 \, ,
\end{cases}
\fe
where as before $q = \exp(2\pi i \tau), \, \bar q = \exp(-2\pi i \bar\tau)$.

Let $\eta$ be a generator of $\bZ_N$.   
First consider the torus partition function $Z^{\eta^r}$ with the insertion of the $\eta^r$ line along the spatial direction:
\ie
\label{TwistBasis}
Z^{\eta^r}(\tau, \bar\tau) &= \sum_{h, \bar h} \left[ n^0_{h, \bar h} + \sum_{Q=1}^{N-1} \omega^{rQ} n^Q_{h, \bar h} \right] \chi_h(\tau) \overline{\chi_{\bar h}(\tau)} \, , \quad \omega = e^{2\pi i \over N} \, ,
\fe
where $n^Q_{h,\bar h} \ge 0$ is the number of  Virasoro primaries with conformal weight $(h,\bar h)$ and $\bZ_N$ charge $Q$ mod $N$.  

The modular S transform of $Z^{\eta^r}(\tau, \bar\tau)$ is a partition function with the insertion of the $\eta^r$ line along the time direction:
\ie
Z_{\eta^r}(\tau, \bar\tau) &=  \sum_{h, \bar h} (n_{\eta^r})_{h, \bar h} \chi_h(\tau) \overline{\chi_{\bar h}(\tau)} \, ,
\fe
where $(n_{\eta^r})_{h, \bar h}$ is the number of Virasoro primaries with conformal weight $(h,\bar h)$ in the defect Hilbert space ${\cal H}_{\eta^r}$. 

To perform bootstrap, it is important to work in a basis where each partition function has an expansion in Virasoro characters with non-negative coefficients.
The twist basis \eqref{TwistBasis} does not have this property, so  we consider instead the charge basis 
\ie
Z^Q (\tau,\bar\tau) &= \text{Tr}_{\substack{|\phi\rangle\in {\cal H}  \\  \widehat\eta|\phi\rangle  = \omega^Q |\phi\rangle}} ~ q^{h-{c/24}} \bar q^{\bar h-{\bar c/24}} 
=  \sum_{h, \bar h} n^Q_{h, \bar h} \chi_h(\tau) \overline{\chi_{\bar h}(\tau)} \, .
\fe
The twist and charge bases are related by a discrete Fourier transform
\ie\label{changebasis}
Z^{\eta^r} (\tau,\bar\tau) =\sum_{Q=0}^{N-1}  \omega^{rQ} Z^{Q}(\tau,\bar\tau) \, .
\fe

\bigskip\bigskip\centerline{\it Bootstrap System}\bigskip

To facilitate the presentation, define the $(2N-1)$-dimensional modular vector ${\bf Z}$ in the charge basis with
\ie
{\bf Z}_i = 
\begin{cases}
Z^{i-1} & \quad i = 1, \dotsc, N \, ,
\\
Z_{\eta^{i-N}} & \quad i = N+1, \dotsc, 2N-1 \, ,
\end{cases}
\fe
and that in the twist basis
\ie
\widetilde{\bf Z}_i = 
\begin{cases}
Z^{\eta^{i-1}} & \quad i = 1, \dotsc, N \, ,
\\
Z_{\eta^{i-N}} & \quad i = N+1, \dotsc, 2N-1 \, .
\end{cases}
\fe
The two are related by the Fourier transform,
\ie
\widetilde{\bf Z} = F \, {\bf Z} \, ,
\fe
where the matrix $F$ can be read off from \eqref{changebasis}.

Since our primary interest is to bound the $\bZ_N$-invariant scalar primary operators, we should work with the partition functions in the charge basis $\bf Z$.  
On the other hand, the modular $S$ matrix is simpler in the twist basis $\tilde {\bf Z}$. 
Let  $\widetilde S$ be the modular $S$ matrix in the twist basis, {\it i.e.} $\tilde{\bf Z}(-1/\tau, -1/\bar\tau) =\widetilde S \,\tilde {\bf Z}(\tau, \bar\tau) $. 
It can be easily written down by noting that $Z^{\eta^r} \leftrightarrow Z_{\eta^r}$ under the S transformation. 
In other words, $\widetilde S$  is the permutation matrix representing the cycle $(2,N+1) (3,N+2) \dotsb (N,2N-1)$.

The modular $S$ matrix in the charge basis is then
\ie
S = F^{-1} \widetilde S \, F \, .
\fe
The bootstrap equation is
\ie
\label{FullBootstrap}
{\bf Z}(-1/\tau, -1/\bar\tau) = S \, {\bf Z}(\tau, \bar\tau) \, .
\fe
Depending on the symmetry and anomaly, this bootstrap equation can be reduced to one of smaller dimensionality.  In Appendix~\ref{Sec:More}, we explain this reduction and present the explicit reduced modular S matrices for small $N$.

The linear functional method for ruling out putative spectra ${\cal P}$ proceeds as in \cite{Lin:2019kpn}.  The idea is to expand the bootstrap equation \eqref{FullBootstrap} into a sum of characters $\chi_h(\tau) \chi_{\bar h}(\bar\tau)$ over the putative spectrum $(h, \bar h) \in {\cal P}$, subject to the appropriate spin selection rule \eqref{SpinSelectionAut}, and search for a linear functional whose action on the combination 
$I \, \chi_h(-1/\tau) \chi_{\bar h}(-1/\bar\tau) - S \, \chi_h(\tau) \chi_{\bar h}(\bar\tau)$ ($I$ is the identity matrix) is of definite sign on the entire ${\cal P}$. If such a functional exists, then the putative spectrum is ruled out. Iterating this procedure produces various constraints, such as bounds on the gap in the spectrum of $\bZ_N$ invariant scalar primaries. 

More specifically, we expand a general linear function $\A$ in the derivative basis, up to a certain derivative order $\Lambda$, such that on a test vector-valued function $f(\tau, \bar\tau)$,
\ie
\A[f] = \sum_{\substack{m=n=0\\m+n \le \Lambda}} \partial_\tau^m \partial_{\bar\tau}^n f(\tau, \bar\tau)|_{\tau = i, \bar\tau = -i} \, .
\fe
The numerical bounds in this paper are all obtained at derivative order $\Lambda = 25$ and with spin truncation $s_\text{max} = 50$.\footnote{The positivity of the linear functional acting on $I \, \chi_h(-1/\tau) \chi_{\bar h}(-1/\bar\tau) - S \, \chi_h(\tau) \chi_{\bar h}(\bar\tau)$ is enforced for $|h - \bar h| \le 50$.
}
The search for a linear function utilizes the semi-definite programming solver SDPB \cite{Simmons-Duffin:2015qma,Landry:2019qug}.\footnote{We use the following SDPB parameter settings: (binary) $\texttt{precision=768}$, \, $\texttt{initialMatrixScalePrimal=1e-10}$, \, $\texttt{initialMatrixScaleDual=1e-10}$, \, $\texttt{maxComplementarity=1e-30}$, \, $\texttt{feasibleCenteringParameter=0.1}$, \, $\texttt{infeasibleCenteringParameter=0.3}$, \, $\texttt{stepLengthReduction=0.7}$.
}

\subsection{Bounds on Symmetry-Preserving Scalar Primaries}\label{sec:numerics}

In the following, we present the numerical bounds on the gap in the spectrum of 
scalar Virasoro primaries for $\bZ_N$ with small $N$ and every possible anomaly.  We will find saturation by several WZW models.  The $c=1$ numerical bounds have to be interpreted differently because there are additional degenerate Virasoro modules at $h=\frac{n^2}{4}$ with $n \in \bZ$, which are not incorporated into the bootstrap system.\footnote{We cannot compare the bounds with \eqref{c=1Gap} since the latter are uncharged scalar gaps for $\mathfrak{u}(1)$ {\it current algebra} primaries. It is a coincidence that for $\bZ_3$ and $\bZ_4$, our bounds agree with the values given by \eqref{c=1Gap}.
}

Note that bootstrap is blind to unfaithfulness: a $\bZ_N$ global symmetry can be  extended to a $\bZ_{pN}$ symmetry realized unfaithfully  for some positive integer $p$.  The discussion in Section \ref{sec:anomaly} implies that our numerical bounds for $\bZ_{pN}$ symmetry with anomaly $k'$ mod $pN$ can be applied to CFTs with a $\bZ_N$ global symmetry that has anomaly $k$ mod $N$ if $k' = p^2 k$ mod $pN$.

\bigskip\bigskip\centerline{\it  No Symmetry}\bigskip

The upper bound on the scalar primaries without assuming any global symmetry was done in \cite{Collier:2016cls}.  
It was found that   relevant operators must be present when the central charge lies within $0 < c < 8$. 
The two ends of this range is saturated by the trivial theory and the $(\mathfrak{e}_8)_1$ WZW model, in which the current bilinear operator is a marginal operator.

\bigskip\bigskip\centerline{\it  $\bZ_2$ Symmetry}\bigskip

Modular bootstrap with $\bZ_2$ global symmetry was studied by the present authors in \cite{Lin:2019kpn}.  
It was found that $\bZ_2$-even relevant/marginal operators must be present when the central charge lies within $1 \le c \le 7$.  Moreover, for anomalous $\bZ_2$  $(k=1)$, the bounds $\Delta_\text{scalar}^{Q=0} = 2$ at $c=1$ and $c=7$ are saturated by the center symmetries of the $\mathfrak{su}(2)_1$ and $(\mathfrak{e}_7)_1$ WZW models, respectively.

\bigskip\bigskip\centerline{\it  $\bZ_3$ Symmetry}\bigskip

The upper bounds on the lightest $\bZ_3$-symmetric scalar operator for different anomalies are shown in Figure~\ref{Fig:Z3Scalar}, and other types of bounds can be found in Appendix~\ref{App:Gap}.  Depending on the anomaly, we find that there must exist a $\bZ_3$-symmetric relevant/marginal scalar primary in the following ranges of the central charge
\ie
\label{Z3Range}
k = 0: \qquad & 1 <  c \le 6 \, , && \qquad \Delta_\text{scalar}^{Q=0} = 2 \text{~~at~~} c = 2, 6 \, ,
\\
k = 1, 2: \qquad & 1 < c < 6.91 \, ,
\\
\text{Any } k: \qquad & 1 < c \le 6 \, .
\fe
For $k=0$, the bounds $\Delta_\text{scalar}^{Q=0} = 2$ at $c=2$ and $c=6$ are saturated by the center symmetries in the $\mathfrak{su}(3)_1$ and $(\mathfrak{e}_6)_1$ WZW models, respectively.\footnote{For $\bZ_3$ symmetry with arbitrary anomaly, 
the bootstrap bound $\Delta_\text{scalar}^{Q=0}$ at $c=1$ coincidentally has the same value as the uncharged scalar gap given in \eqref{c=1Gap} for the $c=1$ compact boson.  However, \eqref{c=1Gap} lists the uncharged scalar gap in the spectrum of exponential operators (\textit{i.e.,} $\mathfrak{u}(1)$ current algebra primaries) and not that in the spectrum of Virasoro primaries.  
The same holds true for $\bZ_4$ with arbitrary anomaly.
\label{c=1foot} 
}

\begin{figure}[H]
\centering
\includegraphics[width=.45\textwidth]{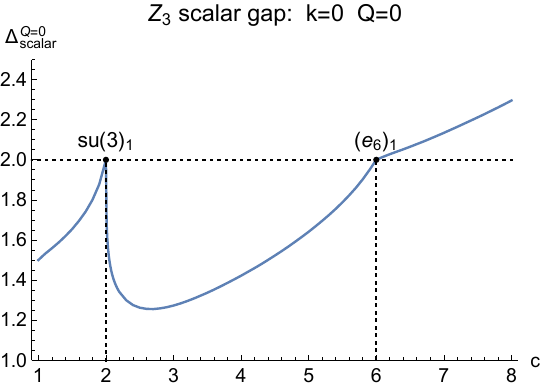} \quad
\includegraphics[width=.45\textwidth]{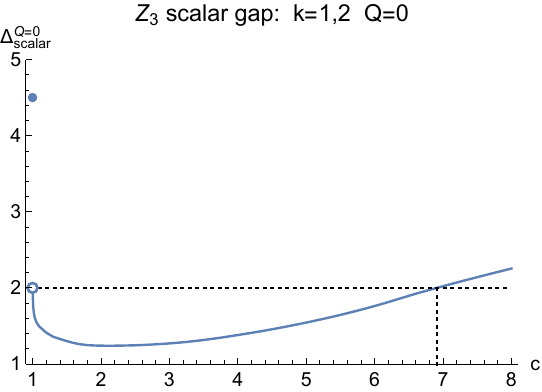}
\caption{Upper bounds  $\Delta_\text{scalar}^{Q=0}$ on the lightest $\bZ_3$-symmetric scalar operator in the spectrum of 
Virasoro primaries with anomaly $k$. 
}
\label{Fig:Z3Scalar}
\end{figure}


\bigskip\bigskip\centerline{\it $\bZ_4$ Symmetry}\bigskip

The upper bounds on the lightest $\bZ_4$-symmetric scalar operator for different anomalies are shown in Figure~\ref{Fig:Z4Scalar}, and other types of bounds can be found in Appendix~\ref{App:Gap}.  We find that there must exist a $\bZ_4$-symmetric relevant/marginal scalar primary in the following ranges of the central charge
\ie
\label{Z4Range}
k = 0: \qquad & 2 \le c \le 6 \, , && \qquad \Delta_\text{scalar}^{Q=0} = 2 \text{~~at~~} c = 2, 4, 6 \, ,
\\
k = 1, 3: \qquad & 2.19 < c < 5.90 \, , 
\\
k = 2: \qquad & 1 < c < 5.74 \, , && \qquad \Delta_\text{scalar}^{Q=0} = 2 \text{~~at~~} c = 3, 5 \, ,
\\
\text{Any } k: \qquad & 2.19 < c < 5.74 \, .
\fe
For $\bZ_4$ with the $k=2$ anomaly, the bounds at $c=1, 3, 5$ are saturated by the $\bZ_4$ center symmetry of the $\mathfrak{so}(2)_1$, $\mathfrak{so}(6)_1$, $\mathfrak{so}(10)_1$ WZW models (see Section~\ref{Sec:SO4N+2}).

Next, we move on to the bounds for a non-anomalous $\bZ_4$ ($k=0$). 
As discussed in Section \ref{sec:anomaly}, a $\bZ_2$ symmetry, anomalous or not, can be  extended to an unfaithful non-anomalous $\bZ_4$ symmetry (see also \cite{Wang:2017loc,Tachikawa:2017gyf}). 
Hence the corresponding $\bZ_2$ twisted partition functions also satisfy or saturate our $\bZ_4$ $k=0$ bootstrap bounds.

 The bounds $\Delta_\text{scalar}^{Q=0} = 2$ at $c=2, 4, 6$ are saturated by a non-anomalous $\bZ_2$ symmetry in the $\mathfrak{so}(4)_1$, $\mathfrak{so}(8)_1$, $\mathfrak{so}(12)_1$ WZW models. 
This $\bZ_2$ symmetry is a center symmetry  discussed in (B.28) of \cite{Lin:2019kpn}.  
At $c = 1$, the bound for non-anomalous $\bZ_4$ is saturated by the $\bZ_4$ subgroup of the momentum ${\rm U}(1)_n$ of the compact boson at radius $R=2$.  The $\bZ_4$-symmetric operators that saturate the bound are ${\cal O}_{4,0}$ and ${\cal O}_{0,1}$ with $\Delta=2$. 
It is also saturated by the anomalous center $\bZ_2$ symmetry, realized unfaithfully as a non-anomalous $\bZ_4$, of the $\mathfrak{su}(2)_1$ WZW model.

The $c_L=c_R={ N\over 2}$ bosonic $\mathfrak{so}(N)_1$ WZW model (sometimes known as the $Spin(N)_1$ WZW model) is the bosonization of $N$ free (non-chiral) Majorana fermions. 
Indeed, the chiral fermion parity $(-1)^{F_L}$ that flips the signs of all left-moving Majorana fermions under bosonization becomes a    $\bZ_2$ symmetry if $N= 0$ or 4 mod 8, and is extended to an anomalous $\bZ_4$ with $k=2$ anomaly if $N= 2$ or 6 mod 8 \cite{Thorngren:2018bhj,Ji:2019ugf}.

\begin{figure}[H]
\centering
\includegraphics[width=.45\textwidth]{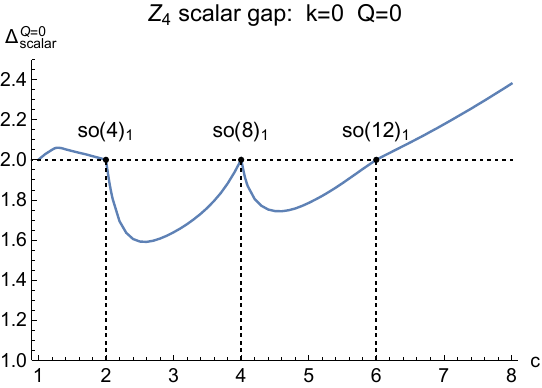} \quad
\includegraphics[width=.45\textwidth]{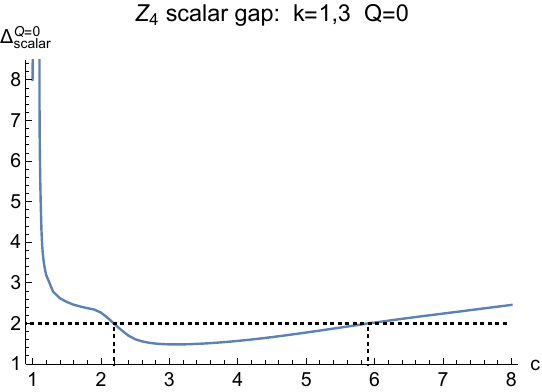}
\\
~
\\
\includegraphics[width=.45\textwidth]{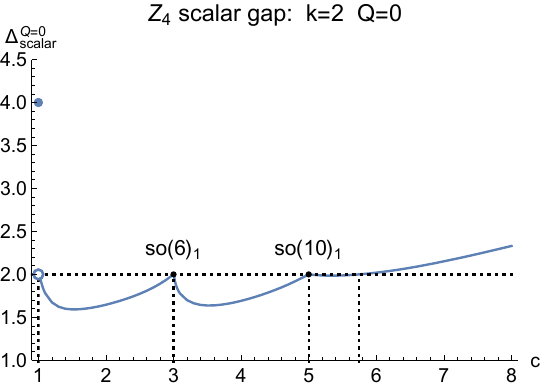} 
~ \hspace{.45\textwidth} ~
\caption{Upper bounds  $\Delta_\text{scalar}^{Q=0}$ on the lightest $\bZ_4$-symmetric scalar operator  in the spectrum of 
Virasoro primaries with anomaly $k$. 
}
\label{Fig:Z4Scalar}
\end{figure}

\bigskip\bigskip\centerline{\it $\bZ_5$ Symmetry}\bigskip

\begin{figure}[H]
\centering
\includegraphics[width=.45\textwidth]{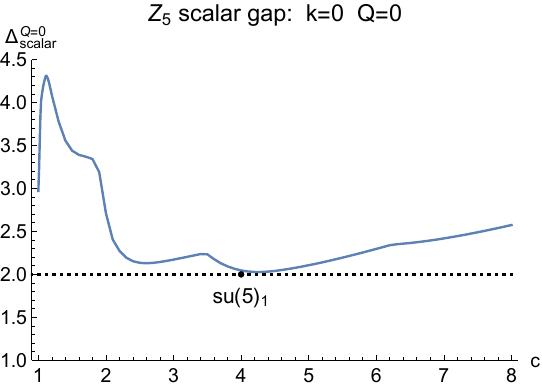} \quad
\includegraphics[width=.45\textwidth]{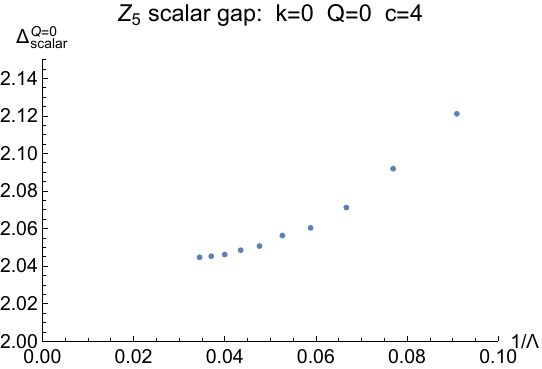}
\\
~
\\
\includegraphics[width=.45\textwidth]{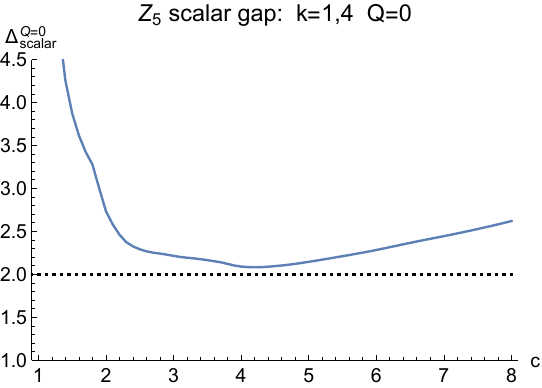}
\includegraphics[width=.45\textwidth]{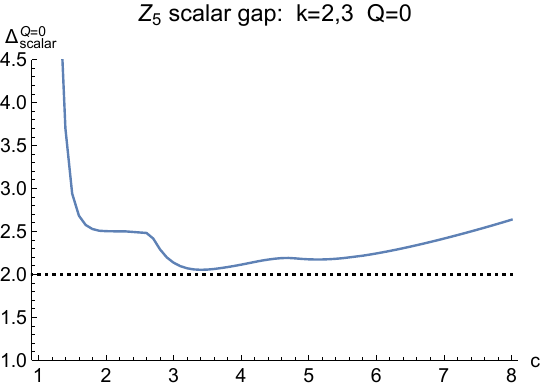}
~ \hspace{.45\textwidth} ~
\caption{Upper bounds $\Delta_\text{scalar}^{Q=0}$ on the lightest $\bZ_5$-symmetric scalar operator  in the spectrum of 
Virasoro primaries with anomaly $k$. The upper-right figure shows the bound over the inverse  derivative order $1/\Lambda$ for non-anomalous $\bZ_5$ at $c = 4$.}
\label{Fig:Z5Scalar}
\end{figure}

For a $\bZ_5$ symmetry with any given 't Hooft anomaly, we do not find any range of the central charge where there must be a $\bZ_5$-symmetric relevant/marginal scalar operator. 
See Figure~\ref{Fig:Z5Scalar}.
For non-anomalous $\bZ_5$ at $c=4$, which is the value of the central charge extrapolated from \eqref{range} to $N=5$, the bound on the scalar primary operator $\Delta_\text{scalar}^{Q=0}  < 2.045$ 
is almost saturated by the marginal current bilinear operator in the $\mathfrak{su}(5)_1$ WZW model.  

Nevertheless, the dependence of the bound on the derivative order $\Lambda$, as shown in the upper-right corner of Figure~\ref{Fig:Z5Scalar}, suggests that the extrapolating to infinite $\Lambda$ does not bring the bound much closer to $\Delta_\text{scalar}^{Q=0} = 2$.  Such ``so close yet so far'' instances concerning bound saturation occur often in numerical bootstrap studies, for instance in the classic (2+1)$d$ Ising bootstrap\cite{ElShowk:2012ht,El-Showk:2014dwa}.
It is often believed that exact saturation will be attained once the full set of bootstrap constraints---beyond modular invariance/covariance and four-point crossing---are taken into account.  While the shrinking of islands \cite{Kos:2014bka} in the (2+1)$d$ Ising case provides some evidence for such belief, it is generally difficult to make such improvements.  Thus we leave open the interpretation of the close-to-saturation by the $\mathfrak{su}(5)_1$ WZW model.

\eject

\bigskip\bigskip\centerline{\it $\bZ_6$ Symmetry}\bigskip

\begin{figure}[H]
\centering
\includegraphics[width=.45\textwidth]{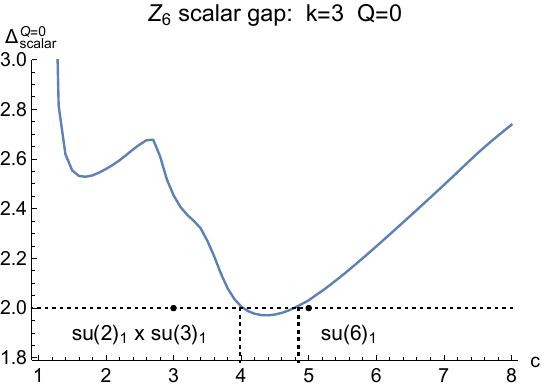}
\caption{Upper bounds  $\Delta_\text{scalar}^{Q=0}$ on the lightest $\bZ_6$-symmetric scalar operator in the spectrum of 
Virasoro primaries with anomaly $k=3$. }
\label{Fig:Z6Scalar}
\end{figure}

The upper bound on the $\bZ_6$-symmetric scalar operator for the $k = 3$ anomaly is shown in Figure~\ref{Fig:Z6Scalar}.  We find that there must exist a $\bZ_6$-symmetric relevant/marginal scalar primary in the range of the central charge
\ie
k = 3: \qquad 3.98 < c < 4.85 \, .
\fe
The bound $\Delta_\text{scalar}^{Q=0} < 2.022$ at $c=5$ is almost saturated by the center symmetry in the $\mathfrak{su}(6)_1$ WZW model.
For other anomalies $k \neq 3$, the bounds $\Delta_\text{scalar}^{Q=0}$ do not drop to or below 2.  Thus if we are blind to the 't Hooft anomaly, then we do not find any range of the central charge where there must be a $\bZ_6$-symmetric relevant/marginal scalar operator.

\bigskip\bigskip\centerline{\it $\bZ_7$ Symmetry}\bigskip

For a $\bZ_7$ symmetry with any given 't Hooft anomaly, we do not find any range of the central charge where there must be a $\bZ_7$-symmetric relevant/marginal scalar operator. 
This presents a natural stopping point for our numerical exploration of constraints on robust gapless fixed points with $\bZ_N$ symmetry.

\section*{Acknowledgement}

We thank Luca Delacr\'etaz, Meng Cheng,  Pranay Gorantla, Theo Johnson-Freyd, Zohar Komargodski, Ho Tat Lam,  Michael Levin, Kantaro Ohmori, and Nathan Seiberg for helpful discussions. 
We are grateful to Meng Cheng, Theo Johnson-Freyd, and Justin Kulp for comments on the first draft.
YL is supported by the Sherman Fairchild Foundation, by the U.S. Department of Energy, Office of Science, Office of High Energy Physics, under Award Number DE-SC0011632, and by the Simons Collaboration Grant on the Non-Perturbative Bootstrap.  SHS is supported by the Simons Collaboration on Ultra-Quantum Matter, which is a grant from the Simons Foundation (651440, NS). 
This research was supported in part by the National Science Foundation under Grant No. NSF PHY-1748958.

\appendix

\section{More on the Modular Bootstrap Equations}
\label{Sec:More}

\bigskip\bigskip\centerline{\it Reduction of the Bootstrap System}\bigskip

We can consider a reduced bootstrap system by combining basis partition functions together.
The reduction takes the schematic form
\ie
{\bf Z}_\text{red} \equiv R \, {\bf Z} \, ,
\fe
where $R$ is the {\it reduction} matrix.  For instance, for $\bZ_N$, if we combine all charge sectors together and all defect partition functions together, then
\ie
\label{RExample}
R =
\begin{pmatrix}
1 & 0 & \dotsb & 0 & 0 & \dotsb & 0
\\
0 & 1 & \dotsb & 1 & 0 & \dotsb & 0
\\
0 & 0 & \dotsb & 0 & 1 & \dotsb & 1
\end{pmatrix} \, .
\fe
The reduced partition vector obeys the crossing equation
\ie
\label{UniversalCrossing}
& {\bf Z}_\text{red}(-1/\tau, -1/\bar\tau) = 
S_\text{red} \, {\bf Z}_\text{red}(\tau, \bar\tau) \, , 
\quad 
S_\text{red} = R \, S R^t \, (R \, R^t)^{-1} \, .
\fe
Given any consistent solution $\bf Z$ to the original bootstrap system \eqref{FullBootstrap}, ${\bf Z}_\text{red} = R \, {\bf Z}$ solves the reduced bootstrap system \eqref{UniversalCrossing}. Hence the constraints from the former cannot be weaker than those from the latter.

Can the constraints from the reduced bootstrap system be as strong as the original one?  If there exists a {\it lifting} matrix $L$ satisfying $L \, S_\text{red} = S \, L$, then any solution ${\bf Z}_\text{red}$ to the reduced bootstrap system can be lifted to a solution ${\bf Z} = L \, {\bf Z}_\text{red}$ of the original bootstrap system.  

\ie
L \, R \, S R^t \, (R \, R^t)^{-1} = S \, L
\fe

In the example \eqref{RExample}, a consistent lifting matrix is given by
\ie
L = R^t (R \, R^t)^{-1} =
\begin{pmatrix}
1 & 0 & 0
\\
0 & \frac{1}{N-1} & 0
\\
0 & \vdots & 0
\\
0 & \frac{1}{N-1} & 0
\\
0 & 0 & \frac{1}{N-1}
\\
0 & \vdots & 0
\\
0 & 0 & \frac{1}{N-1}
\end{pmatrix} \, .
\fe
However, an important caveat is that the lift may not be compatible with the spin selection rule.  To avoid such incompatibility, we should restrict the reduction $R$ to only combine defect partition functions with the same spin selection rule, and require the lift $L$ to respect the spin selection rule.  The example \eqref{RExample} satisfies this restriction for non-anomalous $\bZ_N$ and anomalous $\bZ_{N \le 3}$, but not for anomalous $\bZ_{N \ge 4}$.  If a lift satisfying this restriction exists, then the reduced bootstrap system loses no generality, {\it i.e.} it produces the same bounds as the original; otherwise, it produces (possibly) weaker bounds.

In the following we present the reduced bootstrap systems that lose no generality for $N = 3, 4, 5, 6$.

\bigskip\bigskip\centerline{\it $\bZ_3$ Bootstrap System}\bigskip

Define the partition vector
\ie
{\bf Z}_{\bZ_3} \equiv \begin{pmatrix} Z^0 \\ Z^1 + Z^2 \\ Z_\eta + Z_{\eta^2} \end{pmatrix} \, ,
\fe
which obeys the crossing equation
\ie
\label{Z3Crossing}
{\bf Z}_{\bZ_3}(-1/\tau, -1/\bar\tau) = 
S_{\bZ_3} \, 
{\bf Z}_{\bZ_3}(\tau, \bar\tau) \, , \quad 
S_{\bZ_3} = \left(
\begin{array}{ccccc}
 \frac{1}{3} & \frac{1}{3} & \frac{1}{3} \\
 \frac{2}{3} & \frac{2}{3} & -\frac{1}{3} \\
 2 & -1 & 0 \\
\end{array}
\right) \, .
\fe
The spin selection rule for $Z_\eta + Z_{\eta^2}$ is
\ie
\label{Z3SpinSelection}
h - \bar h \in {k \over 9} + {\bZ \over 3} \, .
\fe

\bigskip\bigskip\centerline{\it $\bZ_4$ Bootstrap System}\bigskip

Define the partition vector
\ie
{\bf Z}_{\bZ_4} \equiv \begin{pmatrix} Z^0 \\ Z^1 + Z^3 \\ Z^2 \\ Z_\eta + Z_{\eta^3} \\ Z_{\eta^2} \end{pmatrix} \, ,
\fe
which obeys the crossing equation
\ie
\label{Z4Crossing}
{\bf Z}_{\bZ_4}(-1/\tau, -1/\bar\tau) = 
S_{\bZ_4} \, 
{\bf Z}_{\bZ_4}(\tau, \bar\tau) \, , \quad 
S_{\bZ_4} = \left(
\begin{array}{ccccc}
 \frac{1}{4} & \frac{1}{4} & \frac{1}{4} & \frac{1}{4} & \frac{1}{4} \\
 \frac{1}{2} & \frac{1}{2} & \frac{1}{2} & 0 & -\frac{1}{2} \\
 \frac{1}{4} & \frac{1}{4} & \frac{1}{4} & -\frac{1}{4} & \frac{1}{4} \\
 2 & 0 & -2 & 0 & 0 \\
 1 & -1 & 1 & 0 & 0 \\
\end{array}
\right) \, .
\fe
The spin selection rule for $Z_\eta + Z_{\eta^3}$ is
\ie
\label{Z4SpinSelection1}
h - \bar h \in {k \over 16} + {\bZ \over 4} \, ,
\fe
and that for $Z_{\eta^2}$ is
\ie
\label{Z4SpinSelection2}
h - \bar h \in {k \over 4} + {\bZ \over 2} \, .
\fe

\bigskip\bigskip\centerline{\it $\bZ_5$ Bootstrap System}\bigskip

Define the partition vector
\ie
{\bf Z}_{\bZ_5} \equiv \begin{pmatrix} Z^0 \\ Z^1 + Z^4 \\ Z^2 + Z^3 \\ Z_\eta + Z_{\eta^4} \\ Z_{\eta^2} + Z_{\eta^3} \end{pmatrix} \, ,
\fe
which obeys the crossing equation
\ie
\label{Z5Crossing}
{\bf Z}_{\bZ_5}(-1/\tau, -1/\bar\tau) = 
S_{\bZ_5} \, 
{\bf Z}_{\bZ_5}(\tau, \bar\tau) \, , \quad 
S_{\bZ_5} = \left(
\begin{array}{ccccc}
 \frac{1}{5} & \frac{1}{5} & \frac{1}{5} & \frac{1}{5} & \frac{1}{5} \\
 \frac{2}{5} & \frac{2}{5} & \frac{2}{5} & \frac{\sqrt{5}-1}{10} & \frac{-1-\sqrt{5}}{10} \\
 \frac{2}{5} & \frac{2}{5} & \frac{2}{5} & \frac{-1-\sqrt{5}}{10} & \frac{\sqrt{5}-1}{10} \\
 2 & \frac{\sqrt{5}-1}{2} & \frac{-1-\sqrt{5}}{2} & 0 & 0 \\
 2 & \frac{-1-\sqrt{5}}{2} & \frac{\sqrt{5}-1}{2} & 0 & 0 \\
\end{array}
\right) \, .
\fe
The spin selection rule for $Z_\eta + Z_{\eta^4}$ is
\ie
\label{Z5SpinSelection1}
h - \bar h \in {k \over 25} + {\bZ \over 5} \, ,
\fe
and that for $Z_{\eta^2} + Z_{\eta^3}$ is
\ie
\label{Z5SpinSelection2}
h - \bar h \in - {k \over 25} + {\bZ \over 5} \, .
\fe

\bigskip\bigskip\centerline{\it $\bZ_6$ Bootstrap System}\bigskip

Define the partition vector
\ie
{\bf Z}_{\bZ_6} \equiv \begin{pmatrix} Z^0 \\ Z^1 + Z^5 \\ Z^2 + Z^4 \\ Z^3 \\ Z_\eta + Z_{\eta^5} \\ Z_{\eta^2} + Z_{\eta^4} \\ Z_{\eta^3} \end{pmatrix} \, ,
\fe
which obeys the crossing equation
\ie
\label{Z5Crossing}
{\bf Z}_{\bZ_6}(-1/\tau, -1/\bar\tau) = 
S_{\bZ_6} \, 
{\bf Z}_{\bZ_6}(\tau, \bar\tau) \, , \quad 
S_{\bZ_6} = \left(
\begin{array}{ccccccc}
 \frac{1}{6} & \frac{1}{6} & \frac{1}{6} & \frac{1}{6} & \frac{1}{6} & \frac{1}{6} & \frac{1}{6} \\
 \frac{1}{3} & \frac{1}{3} & \frac{1}{3} & \frac{1}{3} & \frac{1}{6} & -\frac{1}{6} & -\frac{1}{3} \\
 \frac{1}{3} & \frac{1}{3} & \frac{1}{3} & \frac{1}{3} & -\frac{1}{6} & -\frac{1}{6} & \frac{1}{3} \\
 \frac{1}{6} & \frac{1}{6} & \frac{1}{6} & \frac{1}{6} & -\frac{1}{6} & \frac{1}{6} & -\frac{1}{6} \\
 2 & 1 & -1 & -2 & 0 & 0 & 0 \\
 2 & -1 & -1 & 2 & 0 & 0 & 0 \\
 1 & -1 & 1 & -1 & 0 & 0 & 0 \\
\end{array}
\right) \, .
\fe
The spin selection rule for $Z_\eta + Z_{\eta^5}$ is
\ie
\label{Z6SpinSelection1}
h - \bar h \in {k \over 36} + {\bZ \over 6},
\fe
that for $Z_{\eta^2} + Z_{\eta^4}$ is
\ie
\label{Z6SpinSelection2}
h - \bar h \in {k \over 9} + {\bZ \over 3} \, ,
\fe
and that for $Z_{\eta^3}$ is
\ie
\label{Z6SpinSelection3}
h - \bar h \in {k \over 4} + {\bZ \over 2}.
\fe

\section{Constraints on the Gap of All Primaries}
\label{App:Gap}

\begin{figure}[H]
\centering
\includegraphics[width=.45\textwidth]{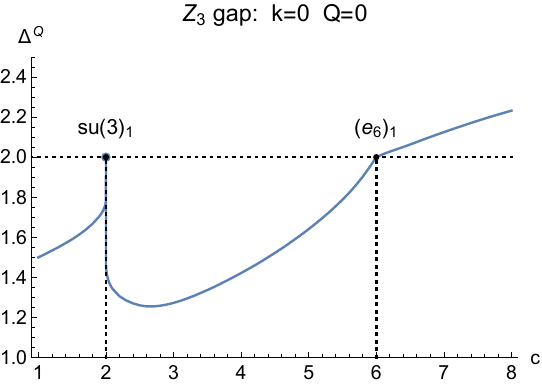} ~ \hspace{.45\textwidth} ~
\\
~
\\
\includegraphics[width=.45\textwidth]{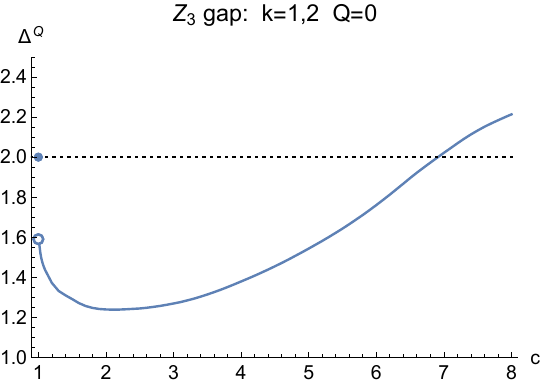} 
\quad
\includegraphics[width=.45\textwidth]{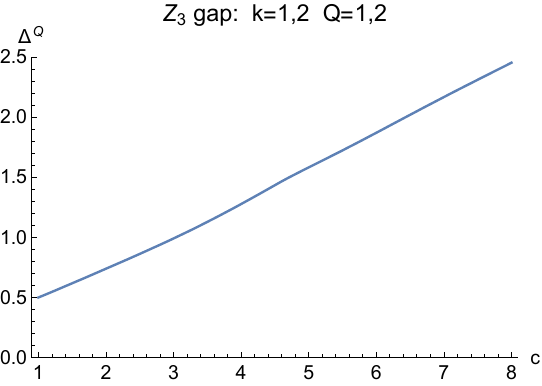}
\caption{Upper bounds on the gap $\Delta^Q$ in the spectrum of 
Virasoro primaries, for $\bZ_3$ anomaly $k = 0, 1, 2$ and $\bZ_3$ charge $Q = 0, 1, 2$.  For $k=0$, $Q=0$, note the discontinuous jump at $c=2$ to $\Delta_\text{scalar}^{Q=0} = 2$.
}
\label{Fig:Z3}
\end{figure}

\begin{figure}[H]
\centering
\includegraphics[width=.45\textwidth]{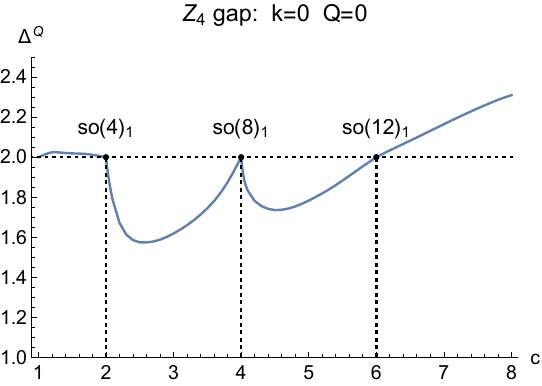} \quad
\includegraphics[width=.45\textwidth]{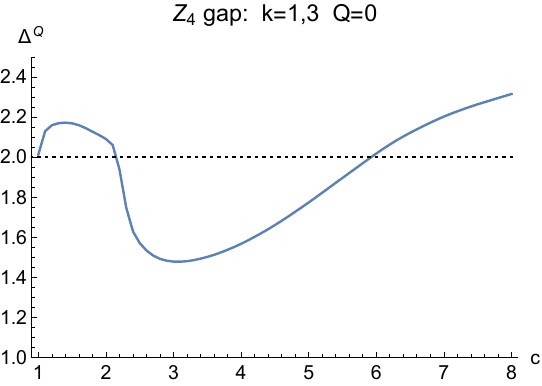}
\\
~
\\
\includegraphics[width=.45\textwidth]{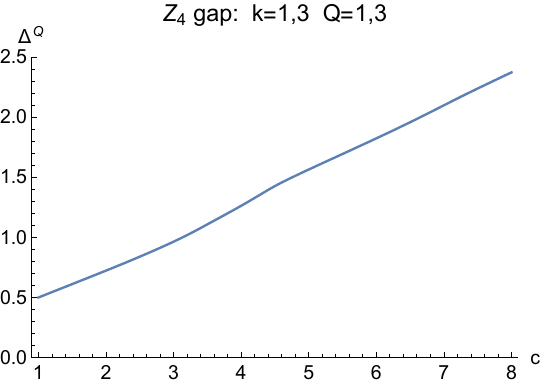} \quad \includegraphics[width=.45\textwidth]{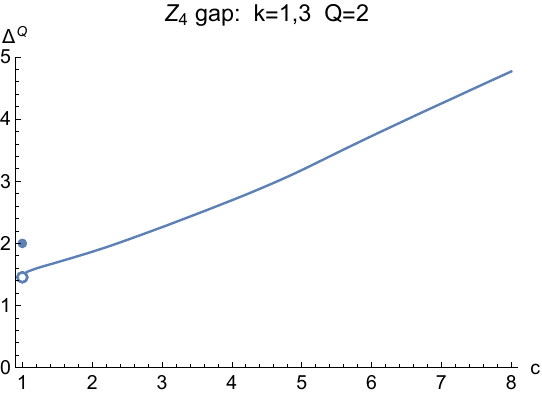}
\\
~
\\
\includegraphics[width=.45\textwidth]{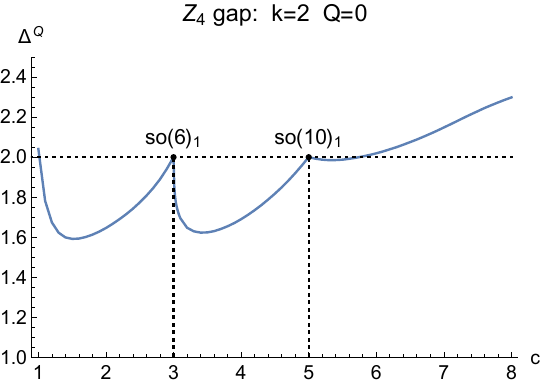} \quad \includegraphics[width=.45\textwidth]{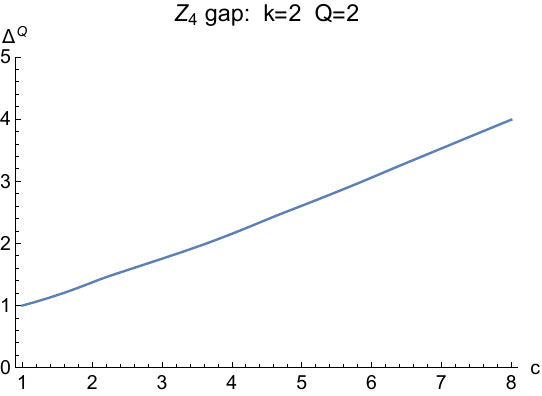}
\caption{Upper bounds on the gap $\Delta^Q$ in the spectrum of 
Virasoro primaries, for $\bZ_4$ anomaly $k = 0, 1, 2, 3$ and $\bZ_4$ charge $Q = 0, 1, 2, 3$.  
}
\label{Fig:Z4}
\end{figure}

\begin{figure}[H]
\centering
\includegraphics[width=.45\textwidth]{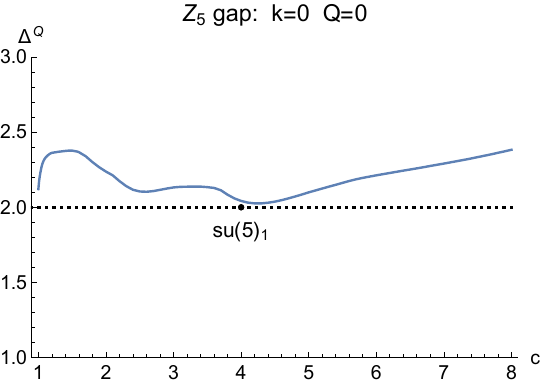} ~ \hspace{.45\textwidth} ~
\\
~
\\
\includegraphics[width=.45\textwidth]{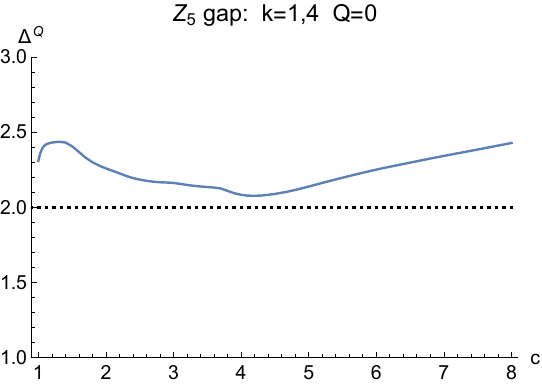} 
\quad
\includegraphics[width=.45\textwidth]{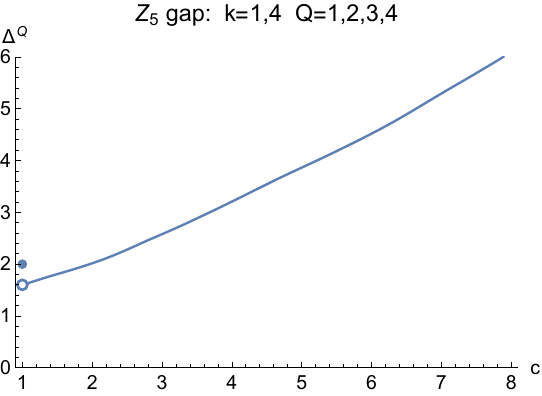}
\\
~
\\
\includegraphics[width=.45\textwidth]{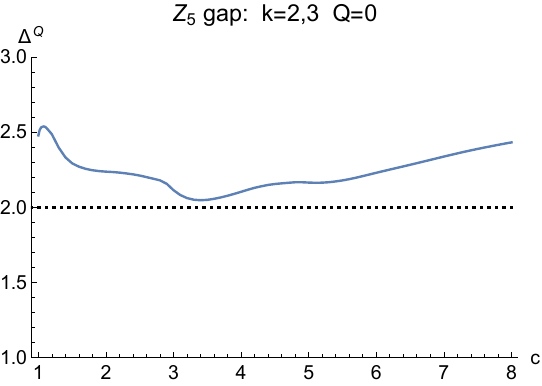} 
\quad
\includegraphics[width=.45\textwidth]{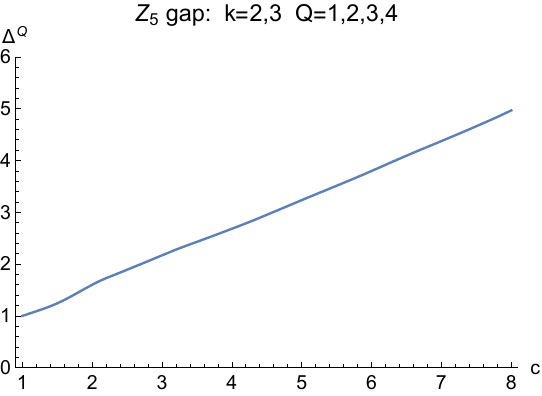}
\caption{Upper bounds on the gap $\Delta^Q$ in the spectrum of 
Virasoro primaries, for $\bZ_5$ anomaly $k = 0, 1, 2, 3, 4$ and $\bZ_5$ charge $Q = 0, 1, 2, 3, 4$.}
\label{Fig:Z5}
\end{figure}

In the main text, we focused on upper bounds on the gap of $\bZ_N$-symmetric scalar primaries since they are relevant for the discussion of robust, symmetry-preserving gapless phases.  However, the general framework of modular bootstrap with symmetry lines can produce more general bounds on  primary operators (not just scalars) with different symmetry charges.  We present these bounds in Figures~\ref{Fig:Z3}, \ref{Fig:Z4} and \ref{Fig:Z5}. 

In the case of a $\bZ_2$ symmetry \cite{Lin:2019kpn}, at a fixed $c$, an upper bound on the scaling dimension of  charged operators only exists if the  symmetry is anomalous. 
Below we see the same phenomenon for more general $\bZ_N$ symmetries. 
Note that there is also no bound  for the $\bZ_4$ symmetry with the $k = 2$  anomaly in the  $Q = 1$ sector, since this is equivalent to bounding operators charged (odd) under its non-anomalous $\bZ_2$ subgroup.

~\\

\bibliographystyle{JHEP}
\bibliography{zn}

\end{document}